\documentclass[11pt]{article} 
\usepackage[top=1in, bottom=1in, left=1in, right=1in]{geometry}
\usepackage{verbatim,amssymb,epsfig}  
\usepackage{authblk}
\usepackage{amsmath, amsfonts, amstext,amsthm}
\usepackage{enumerate}
\usepackage{csquotes}
\usepackage{enumitem}
\usepackage{natbib}
\usepackage{graphicx}
\usepackage{subcaption}
\usepackage{booktabs} 
\usepackage{hyperref}
\hypersetup{
colorlinks = true,
allcolors = blue
}

\usepackage{url} 
\usepackage[capitalize, nameinlink]{cleveref}
\usepackage{colortbl} 
\usepackage[table]{xcolor} 
\theoremstyle{plain}
\newtheorem{theorem}{Theorem}[section]

\newtheorem{proposition}{Proposition}[section]

\theoremstyle{definition}
\newtheorem{definition}{Definition}[section]

\theoremstyle{remark}
\newtheorem{remark}{Remark}[section]

\theoremstyle{definition}
\newtheorem{example}{Example}[section]

\usepackage{multirow, makecell}

\title{Loss-based Objective and Penalizing Priors for Model Selection Problems}  
\author{Changwoo J. Lee \thanks{Department of Statistics, Texas A\&M University. \texttt{c.lee@stat.tamu.edu}}}
\date{November 2023}

\usepackage{bm}
\usepackage{diagbox}

\usepackage{macros}

\begin{document}

\maketitle

\begin{abstract}

Many Bayesian model selection problems, such as variable selection or cluster analysis, start by setting prior model probabilities on a structured model space. 
Based on a chosen loss function between models, model selection is often performed with a Bayes estimator that minimizes the posterior expected loss. 
The prior model probabilities and the choice of loss both highly affect the model selection results, especially for data with small sample sizes, and their proper calibration and careful reflection of no prior model preference are crucial in objective Bayesian analysis. 
We propose \textit{risk equilibrium priors} as an objective choice for prior model probabilities that only depend on the model space and the choice of loss. Under the risk equilibrium priors, the Bayes action becomes indifferent before observing data, and the family of the risk equilibrium priors includes existing popular objective priors in Bayesian variable selection problems. We generalize the result to the elicitation of objective priors for Bayesian cluster analysis with Binder's loss. 
We also propose \textit{risk penalization priors}, where the Bayes action chooses the simplest model before seeing data. 
The concept of risk equilibrium and penalization priors allows us to interpret prior properties in light of the effect of loss functions, and also provides new insight into the sensitivity of Bayes estimators under the same prior but different loss. We illustrate the proposed concepts with variable selection simulation studies and cluster analysis on a galaxy dataset.

\end{abstract}
\begin{keywords}
Objective Bayes, loss function, Bayesian model selection, Bayesian variable selection, Bayesian cluster analysis, random partition priors.
\end{keywords}

\section{Introduction}

Bayesian model selection problems arise from many different contexts of scientific problems, such as Bayesian variable selection \citep[BVS;][]{Tadesse2021-zs}, Bayesian cluster analysis \citep[BCA;][]{Wade2023-oc}, and structure learning problems \citep{Koller2009-qz}. 
The Bayesian model selection procedure begins with the elicitation of prior model probabilities $\pi(M)$. Then, based on the priors of model-specific parameters and the resulting marginal likelihood $f(D\given M)$, 
the posterior model probabilities $\Pi(M\given D)$ or their ratios are obtained. While there are many ways to summarize posterior model probabilities into a single model, one of the principled model selection methods is to use a Bayes estimator with a choice of loss function, which selects the model that minimizes posterior expected loss. A popular example is the median probability model in BVS \citep{Barbieri2004-ux}, which selects variables that have a marginal inclusion probability greater than 0.5, and it is the Bayes estimator under the Hamming loss \citep{Carvalho2008-ip}. The Bayes estimator has a solid decision-theoretic background and possesses many attractive properties \citep{Berger1985-qi, Robert2007-mo}; we focus on model selection procedures based on a Bayes estimator.

\begin{figure}
    \centering
    \includegraphics[width = 0.9\textwidth]{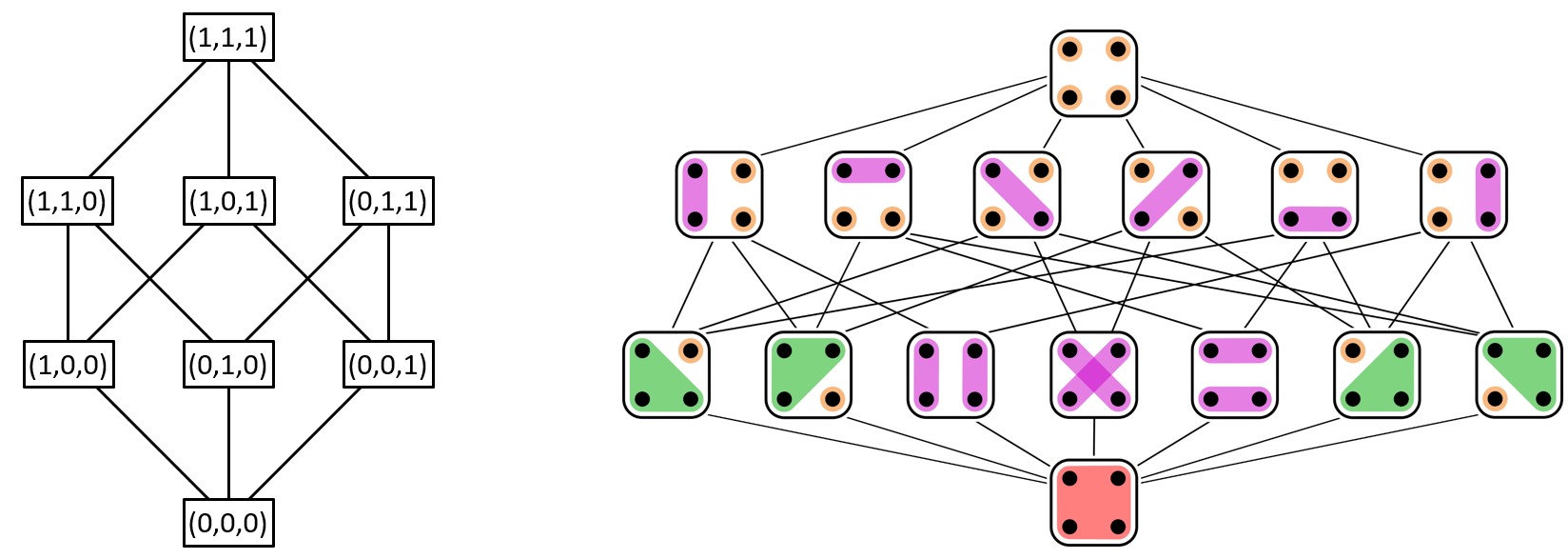}
    \caption{Bayesian model selection problem examples with a structured model space $\calM$. Edges between models describe a partial order relation that compares model complexity, with a simpler model placed at bottom. (Left) $\calM = \{0,1\}^3$ in Bayesian variable selection (BVS) problem with $3$ variables. (Right) $\calM = \{M: M \text{ is a partition of }\{1,2,3,4\}\}$ in Bayesian cluster analysis (BCA).}
    \label{fig:modelsp_examples}
\end{figure}

We consider model selection problems where the model space is finite, structured, and equipped with a partial order that compares model complexity. This characteristic is highly common in many scenarios, encompassing all previous examples of BVS, BCA, and structure learning; see Figure~\ref{fig:modelsp_examples} for an illustration and also see \citet{Taeb2023-ed} for more examples. The structured model space allows us to explore various concepts of ``dissimilarity'' between models, encoded as a loss function $L(M, \hat{M})$ that quantifies the ``cost'' of choosing the model $\hat{M}$ when the true is $M$. 

The prior model probabilities $\pi$ have a crucial impact on model selection results, and a natural question is what are the candidates of $\pi$ when we have no prior preferences on models, but have a loss function $L$ in mind, and desire to elicit $\pi$ in an ``objective'' manner. 
When the model space is unstructured, uniform prior would be the only sensible choice \citep{Berger2001-nc,Berger2012-iq}, but the structural information of the model space opens up a wide possibility of interpretation of being ``objective'' and corresponding nontrivial objective priors.

The development of objective Bayes methods for model selection problems has only started recently, and the underlying conceptual framework to build objective priors for model selection problems is highly different from the ones for estimation and prediction; see \citet{Consonni2018-iy} for a comprehensive review and references therein. 
However, a growing body of literature mainly focuses on the elicitation of model-specific parameters \citep{Berger2001-nc, Liang2008-ue, Bayarri2008-ri, Bayarri2012-ho,Simpson2017-vc}, and relatively little attention has been devoted to the elicitation of objective prior model probabilities, partly due to the discrete nature of the model space which is completely different from a continuous space (e.g. defining gradient or Hessian is highly nontrivial). 
For BVS, \citet{Cui2008-iu,Ley2009-ur, Scott2010-rp} suggested priors that provide multiplicity control, including the one that induces a uniform distribution on model sizes. \cite{Villa2015-ge} considered a notion of ``worth'' of selecting models based on Kullback-Leibler (KL) divergence between data likelihoods, and \cite{Villa2020-bv} introduced an additional term that penalizes more complex models. However, all previous examples are not easily generalizable to settings other than BVS, or involve integration over the minimum of KL divergence, which could be complex to derive. 

\begin{figure}
    \begin{subfigure}{\textwidth}
        \centering
        \includegraphics[width=0.8\linewidth]{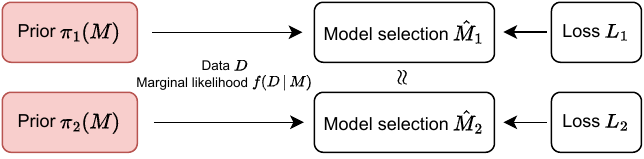}
    \caption{Q1. How can we elicit ``objective'' model priors $\pi(M)$ for a given loss $L$? In particular, how can we make model selection results become less sensitive to the choice of loss, especially when the sample size is small? \vspace{5mm}}
    \end{subfigure}
    \begin{subfigure}{\textwidth}
        \centering
        \includegraphics[width=0.8\linewidth]{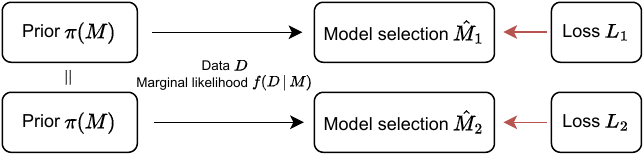}
    \caption{Q2. How can we better understand loss functions’ effect on Bayes estimators? That is, why does a loss $L_1$ give a more parsimonious model but another loss $L_2$ does not?}
    \end{subfigure}
    \caption{Two key research questions explored in this paper.}
    \label{fig:keyq}
\end{figure}

Motivated by the Bayes estimator that chooses the model that minimizes the expected loss, we propose a new family of ``objective'' model priors that is associated with a given loss function, called \textit{risk equilibrium priors}. 
Specifically, we consider a family of prior distributions on models that leads to the Bayes action become neutral before observing data. 
We note that the choice of loss is inherently subjective (see \S 1.6.5 of \citet{Berger1985-qi}) and we do not attempt to ``objectively'' choose a loss, but rather define ``objectivity'' relative to the choice of the loss function. This is similar to the construction of objective priors that are defined relative to a given statistical model; see \citet[\S 3.2]{Berger2006-xo} for more discussion. The proposed concept of ``objectivity'' associated with a loss function can be also understood as an attempt to calibrate Bayesian model selection results across different choices of loss functions. This calibration is particularly crucial for problems with small sample sizes, since prior model probabilities strongly influence posterior and thus model selection results. See Figure~\ref{fig:keyq}(a) for an illustration.

The concept of risk equilibrium priors associated with a loss allows us to easily generalize the proposed idea to model selection problems beyond BVS. We pay particular attention to Bayesian cluster analysis (BCA), where the model space is a collection of partitions and random partition models \citep{Muller2013-yc} serve as model priors. We illustrate how ``objective'' priors for Bayesian cluster analysis can be chosen based on a popular loss function called Binder's loss \citep{Binder1978-cr}. In Section 3.3, we provide concrete examples of Bayesian clustering methods that correspond to the proposed ``objective'' priors, which could be useful for practitioners.


We also define a family of prior distributions on models associated with a loss that leads to the Bayes action choosing the simplest model before seeing data, called \textit{risk penalization priors}. 
The penalized (or sparsity-inducing) priors that are common in high-dimensional model selection problems typically belong to this family with a popular choice of loss function. 
Especially, motivated by the phenomenon that Bayesian cluster analysis results can vary significantly with different choices of loss functions but with the same prior \citep{Wade2018-mp}, we provide a new perspective on the interpretation of loss functions' effect on Bayes estimators by comparing loss-based objective and penalized priors, as described in Figure~\ref{fig:keyq}(b).

To better support our claim and address two key research questions explored in this paper (see Figure~\ref{fig:keyq}), we conclude the paper with performing BVS simulation studies and conducting BCA real data analysis with galaxy data, a popular real benchmark dataset used for cluster analysis.

\section{Review of Bayesian model selection problems}

\subsection{Notations and basic concepts}

Let $M_1,M_2,\dots$ be models and $\calM$ be a discrete, finite space of models. We consider model spaces that are naturally equipped with a partial order relation $\prec$ that compares the complexity of models. That is, $M_1 \prec M_2$ implies that $M_1$ and $M_2$ are comparable and model $M_2$ is more complex than model $M_1$. Here, a model being more complex generally refers to the case when a model has more parameters. Again, figure~\ref{fig:modelsp_examples} shows the structures of $\calM$ and the partial orders in the BVS and BCA. We use the generic notation $p$ to represent a quantity that determines the size of the model space; for example, $p$ is the number of variables in BVS and $p$ is the number of data points to be clustered in BCA.

The model selection problem aims to select the model among $\calM$ that best describes the data $D$. 
A Bayesian approach starts from the prior on the model space $\pi$ and updates the belief based on the marginal likelihood $f( D\given M) = \int f(D\given \vartheta, M)p(\vartheta \given M) d\vartheta$, where the model parameters $\vartheta$ are integrated out. Throughout the paper, we assume proper priors on the model parameters $p(\vartheta\given M)$. It yields the posterior
\begin{equation}
\label{eq:bayesianmodelselectpost}
\Pi(M\given D) = \frac{\pi(M)f(D\given M)}{\sum_{M\in \calM} \pi(M)f(D\given M)}
\end{equation}
and the Bayesian model selection problem seeks to find the best model $\hat{M}$ using the information in the posterior distribution \eqref{eq:bayesianmodelselectpost}. 
We focus on model selection procedure with Bayes estimator, where given a loss function $L(M,\hat{M})$ that quantifies the ``cost'' associated with choosing $\hat{M}$ (action) when the true model is $M$, the Bayes estimator is the minimizer of posterior expected loss 
\begin{equation}
    \hat{M}^{\mathrm{Bayes}} = \argmin_{\hat{M}\in \calM} \bbE^{\Pi}[L(M, \hat{M})] = \argmin_{\hat{M}\in \calM} \sum_{M\in \calM}L(M,\hat{M})\Pi(M \given D).
\end{equation}
Here, the expected loss is called risk $R^\Pi(\hat{M}) = \bbE^{\Pi}[L(M, \hat{M})]$.  
The loss function $L:\calM\times \calM\to [0,\infty)$ need not be symmetric, but throughout the paper we assume that $L$ is bounded below by 0 and $L(M,\hat{M}) = 0$ if and only if $M=\hat{M}$ (identity of indiscernibles). If $L$ is symmetric and satisfies triangular inequality, we also use the term distance interchangeably. 
Bayes estimators encompass a variety of ways to perform model selection based on different loss functions. For example, under the zero-one loss $L_{01}(M, \hat{M}) = \ind (M\neq \hat{M})$, the Bayes estimator corresponds to the model that maximizes the posterior, also called posterior mode or the highest probability model. Other choice of loss leads to different Bayes estimators; see examples in the following sections.

\subsection{Bayesian variable selection (BVS)}

Bayesian variable selection (BVS) problem seeks to characterize the subset of $p$ predictors that best describes the data; see \cite{Tadesse2021-zs} and references therein. The model space of BVS is an hypercube $\calM = \{0,1\}^p$ with cardinality $2^p$, where each model is represented as a binary vector $\bm\gamma =(\gamma_1,\dots,\gamma_p)\in\{0,1\}^p = \calM$ where $\gamma_i = 1$ indicates $i$th predictor is included in the model. 
The partial order that compares model complexity is defined as an inclusion relationship of nonzero indices; we write $\bm\gamma \prec \bm\gamma'$ if $\{i: \gamma_i\neq 0\} \subsetneq \{i': \gamma_i' \neq 0\}$. 
Its covering pairs $(\bm\gamma, \bm\gamma')$, when there is no $\bm\gamma''$ such that $\bm\gamma\prec \bm\gamma'' \prec \bm\gamma'$, corresponds to the edge of the hypercube $\calM$ and the null model $\bm\gamma = \bm{0}_p$ is the least element with respect to $\prec$. 

The zero-one loss and Hamming loss are two popular loss functions used in the BVS literature,
\[
L_{01}(\bm\gamma,\hat{\bm\gamma}) = \ind(\bm\gamma \neq  \hat{\bm\gamma}) \quad \text{(zero-one loss)},\qquad L_{\mathrm{H}}(\bm\gamma,\hat{\bm\gamma}) = \sum_{i=1}^p \ind(\gamma_i \neq \hat{\gamma_i}) \quad \text{(Hamming loss)},
\]
and we also introduce the generalized Hamming loss, an asymmetric version of Hamming loss $L_{\mathrm{GH}(a)}$ with weight $a\in(0,2)$ that puts different costs associated with false positives and false negatives. 

\begin{definition}[Generalized Hamming loss] Consider the BVS model space $\calM = \{0,1\}^p$. We say $L_{\mathrm{GH}(a)}$ is a generalized Hamming loss with weight $a\in(0,2)$ if
\begin{equation}
\label{eq:ghloss}
   L_{\mathrm{GH}(a)}(\bm\gamma,\hat{\bm\gamma}) = \sum_{i=1}^p \left( a \ind(\gamma_i = 0)\ind(\hat{\gamma_i} = 1) +  (2-a) \ind(\gamma_i = 1)\ind(\hat{\gamma}_i = 0) \right) \quad \text{(GH$(a)$ loss)}
\end{equation}
which reduces to (symmetric) Hamming loss when $a=1$.
\end{definition}

\begin{remark}[Median and quantile probability model] The highest probability model, the posterior mode, is the Bayes estimator under the zero-one loss.
Although the highest probability model might sound like a natural choice, \citet{Barbieri2004-ux} showed that the highest probability model is not always an optimal model in terms of predictive performance, and proposed median probability model that chooses variables whose marginal inclusion probability is greater than 0.5. This often differs from the highest probability model; see also \citet{Barbieri2021-er} for additional properties. 
One less well-recognized fact in the literature is that the median probability model is the Bayes estimator under the Hamming loss \citep{Carvalho2008-ip}, which provides a decision-theoretic interpretation of the median probability model. As a natural extension of the median probability model, the quantile probability model \citep{Heyard2019-cz} was proposed that chooses variables whose marginal probability is greater than some threshold other than 0.5. 
However, its connection to the Bayes estimator has been previously overlooked, and we show that the quantile probability model is the Bayes estimator under the generalized Hamming loss.
\end{remark}

\begin{proposition}
\label{prop:ghloss_riskeq}
Consider BVS model space $\calM = \{0,1\}^p$.  The Bayes estimator under the generalized Hamming loss $L_{\mathrm{GH}(a)}$ is $(0.5a)$-quantile probability model, the model consisting of variables whose marginal inclusion probability $\bbP^\Pi(\gamma_i =1)$ is greater than $0.5a$. 
\end{proposition}

We defer all proofs of Propositions and Theorems to Appendix~\ref{proof:proofs}. The above proposition provides a decision-theoretic justification of the quantile probability model. When we choose generalized Hamming loss with $1<a<2$, the selection threshold becomes higher and the resulting Bayes estimator becomes more parsimonious (i.e., contains fewer number of variables) compared to the median probability model based on the symmetric Hamming loss. Similarly when $0<a<1$, the resulting Bayes estimator tends to contain more variables compared to the median probability model, due to less penalization on the false positives than false negatives.

\subsection{Bayesian cluster analysis (BCA)}

Bayesian cluster analysis (BCA) aims to find a partition of data points that best describes the data with some notion of similarity, where examples include mixture models, change point detection models, and stochastic blockmodels; see \cite{Wade2023-oc} for a recent thorough review. Here we consider clustering problems where the number of clusters is unknown. The BCA model space is a partition space $\calM = \{\bm\rho: \bm\rho\text{ is a partition of }\{1,\dots,p\}\}$, denoted as $\calP_{[p]}$ following the notation of \cite{Pitman2006-zf}. Its cardinality is the $p$th Bell number $\scrB_p$ which grows faster than exponentially in terms of $p$. 
The partial order that compares model complexity is the set refinement relation, where we write $\bm\rho \prec \bm\rho'$ if any element (cluster) of $\bm\rho'$ is a subset of some element of $\bm\rho$ and $\bm\rho'\neq \bm\rho$. For example, $\bm\rho = \{\{1,2\},\{3,4\}\} \prec \{\{1\},\{2\}, \{3,4\}\}= \bm\rho'$. The one-block partition is the least element with respect to $\prec$, representing the simplest model in BCA. 
In addition to the set partition notation, we use cluster label representation interchangeably (although the representation is not unique); we denote the partition with cluster labels $\bfz = (z_1,\dots,z_p)\in \{1,\dots,|\bm\rho|\}^p$ where $z_i=j$ if $i$th data point is in the $j$th cluster. For example, $z_1 = z_2$ indicates that the first and second data points are in the same cluster.

We review two popular loss functions in BCA, Binder's loss and VI loss. First is Binder's loss \citep{Binder1978-cr}, where its symmetric version has a one-to-one connection with the Rand index \citep{Rand1971-ic} and the Mirkin metric \citep{Mirkin1996-xc} that is widely used for comparing clusterings. Below, we provide a definition of generalized Binder's loss that allows asymmetry, in fact, the version that Binder originally proposed. 
\begin{definition}[Generalized Binder's loss] Consider the BCA model space $\calM$, a set of partitions of $\{1,\dots,p\}$. We say $L_{\mathrm{GB}(a)}$ is a generalized Binder's loss with weight $a\in (0,2)$ if
    \begin{equation}
\label{eq:gbinderloss}
L_{\mathrm{GB}(a)}(\bfz, \hat\bfz) = \sum_{1\le i<j\le p}\left(a\ind(z_{i}=z_{j}) \ind(\hat{z}_{i} \neq \hat{z}_{j})+ (2-a)\ind(z_{i} \neq z_{j}) \ind(\hat{z}_i=\hat{z}_j)\right) \quad \text{(GB$(a)$ loss)}.
\end{equation}
\end{definition}

When $a=1$, we simply refer to it as Binder's loss and denote it by $L_{\mathrm{B}}$. When $a\neq 1$, it places different costs on failing to separate two units which should be together and failing to group two units which should be separate.
The asymmetric loss function can mitigate the over-clustering behavior of the Bayes estimator, see \cite{Dahl2022-qd} and also \citet{Dombowsky2023-jj}.

The Bayes estimator under the generalized Binder's loss was studied in several research works on Bayesian clustering \citep{Lau2007-ta,Dahl2007-tz,Dahl2022-qd}. When $1<a<2$, since it heavily penalizes the pairs that are separate but should be together, the resulting Bayes estimator leads to a more parsimonious model (i.e. has less number of clusters) compared to the Bayes estimator under the symmetric Binder's loss. The Bayes estimator under the GB(a) loss can be written as the maximizer of $f(\hat{\bfz}) = \sum_{i<j} \ind(\hat{z}_i = \hat{z}_j)(C_{ij} - (2-a)/2)$ \citep{Lau2007-ta}, where the posterior is summarized in terms of posterior co-clustering probabilities $C_{ij} = \bbP^\Pi(z_i = z_i)$. Rather than writing the Bayes estimator as the maximizer of $f(\hat{\bfz})$ involving summation over selected pairs $(i,j)$ based on $\hat{\bfz}$, we provide a new simple, intuitive representation of the Bayes estimator under the GB(a) loss using the check function $\rho_\tau(x) = 2x(\tau-\ind(x<0))$ with $\tau\in(0,1)$.

\begin{proposition}
\label{prop:gbloss_riskeq}
Consider BCA model space $\calM = \calP_{[p]}$ and denote $C_{ij} = \bbP^\Pi(z_i = z_j)$ be posterior co-clustering probabilities for $i\neq j$. The Bayes estimator under the GB(a) loss is the minimizer of the sum of difference between co-clustering probabilities that are transformed by a check function,
\begin{equation}
\label{eq:gbbayesest}
    \hat{\bfz}^{\mathrm{Bayes}} = \argmin_{\hat{\bfz}} R^\Pi_{\mathrm{GB}(a)} =  \argmin_{\hat{\bfz}} \sum_{1\le i<j \le p} \rho_{(2-a)/2}\left(\ind(\hat{z}_i = \hat{z}_j)-C_{ij}\right).
\end{equation}
\end{proposition}

Next, we review another popular loss called the variation of information (VI) loss \citep{Meila2007-co}, which is closely related to the concept of mutual information and also can be axiomatically defined \citep{Meila2005-dx}. The VI loss can be written in terms of cluster label representation,
\begin{align}
L_{\mathrm{V}}(\mathbf{z}, \hat{\mathbf{z}}) &= \frac{1}{p}\sum_{i=1}^p \log_2 \left(\sum_{j=1}^p \ind(z_i=z_j)\right) + \frac{1}{p}\sum_{i=1}^p \log_2 \left(\sum_{j=1}^p \ind(\hat{z}_i=\hat{z}_j)\right) \nonumber\\
 &- \frac{2}{p}\sum_{i=1}^{p} \log_2\left(\sum_{j=1}^p \ind(z_i = z_j)\ind(\hat{z}_i = \hat{z}_j) \right) \quad \text{(VI loss)},\label{eq:viloss}
\end{align}
and also see \cite{Dahl2022-qd} for the generalized version with unequal weights. The Bayes estimator under the VI loss minimizes the posterior expected VI loss, called posterior VI risk,
\begin{equation}
\label{eq:virisk}    
R_V^\Pi(\hat{\bfz}) = H + \frac{1}{p}\sum_{i=1}^p \log_2 \left(\sum_{j=1}^p \ind(\hat{z}_i=\hat{z}_j)\right) - \frac{2}{p}\sum_{i=1}^p\bbE^\Pi\left[\log_2 \left(\sum_{j=1}^p \ind (z_i = z_j)\ind(\hat{z}_i=\hat{z}_j)\right)\right],
\end{equation}
where $H = \bbE^\Pi\left[p^{-1}\sum_{i=1}^p \log_2 \sum_{j=1}^p \ind(z_i=z_j)\right]$ is a constant. Since \eqref{eq:virisk} contains an expectation over a nonlinear term that depends on both $\Pi$ and $\hat{\bfz}$ that increases the computational burden of the calculation of Bayes estimator, \citet{Wade2018-mp} proposed to approximate the VI risk \eqref{eq:virisk} with its lower bound obtained from Jensen's inequality. That is, approximate $R^\Pi_V(\hat{\bfz})$ with 
\begin{equation}
\label{eq:virisklb}
\tilde{R}_V^\Pi(\hat{\bfz}) = H + \frac{1}{p}\sum_{i=1}^p \log_2 \left(\sum_{j=1}^p \ind(\hat{z}_i=\hat{z}_j)\right) - \frac{2}{p}\sum_{i=1}^p\log_2 \left(\sum_{j=1}^p C_{ij}\ind(\hat{z}_i=\hat{z}_j)\right) \le R_V^\Pi(\hat{\bfz})
\end{equation}
so that $\tilde{R}_V^\Pi(\hat{\bfz})$ depends on posterior $\Pi$ only through co-clustering probabilities $C_{ij}$. 

\begin{remark}[Finding Bayes estimator in BCA]
Unlike BVS, finding the minimizer of the posterior risk in BCA is a nontrivial task, since combinations of pairs and non-pairs that minimize the risk for each summand may not lead to a valid partition. Existing methods include using hierarchical clustering dendrograms \citep{Medvedovic2002-tt,Fritsch2009-tl}, search within the posterior samples \citep{Dahl2006-ym}, binary integer programming \citep{Lau2007-ta}, greedy algorithms \citep{Wade2018-mp,Rastelli2018-jq}, and the most recently proposed sequentially-allocated latent structure optimization (SALSO) algorithm \citep{Dahl2022-qd}.
\end{remark}

\section{Risk equilibrium priors with respect to a loss}

\subsection{Definition and role as an objective priors}

We introduce \textit{risk equilibrium priors} based on the behavior of \textit{prior} expected loss, motivated by the Bayesian decision rule that decides the model that minimizes the posterior expected loss. 

\begin{definition}[Risk equilibrium prior] We say a prior $\pi$ is a \textit{risk equilibrium prior with respect to loss $L$, or with respect to risk $R^\pi$}, when the prior risk $R^\pi(\cdot) = \bbE^{\pi}[L(M,\cdot)]$ becomes a constant function over $\calM$.
\end{definition}

The proposed risk equilibrium prior does not uniquely determine a probability distribution; rather, it characterizes a family of priors that have certain properties with respect to a loss function. Risk equilibrium priors can be considered as a family of objective priors that depends on the choice of the loss function, in the sense that the Bayes action is indifferent before seeing the data. It parallels the idea of \citet{Leisen2020-sd} who proposed objective priors derived from a scoring rule being a constant for the continuous parameter space. Most importantly, the characterization of the risk equilibrium prior does not depend on the data likelihood, and borrowing words from \cite{Leisen2020-sd}, risk equilibrium priors could also be argued as ``more objective'' than those depend on data likelihood; e.g. \citet{Villa2015-ge}.

We provide a careful interpretation of risk equilibrium priors as a constant scoring rule following the rationale of \citet{Leisen2020-sd}. Scoring rules \citep{Gneiting2007-jx} quantify the quality of probabilistic forecasts given the realization, which can be thought of as loss functions between probabilistic forecasts and the value that materializes. 
\citet{Leisen2020-sd} proposed a class of objective priors supported on a \textit{continuous} parameter space that arises from setting a proper scoring rule $S(\pi,\theta)$ being a constant. Here scoring rule $S$ is called \textit{proper relative to} $\calP$ when $S$ satisfies $S(\nu, \nu)\le S(\pi,\nu)$ for all $\pi,\nu\in \calP$, where we write $S(\pi,\nu) = \int_\Theta S(\pi,\theta) d \nu(\theta)$, and called \textit{strictly proper} when equality holds if and only if $\pi = \nu$. \citet{Leisen2020-sd} considered an example when the scoring rule is set as a combination of the log score and the Hyv\"arinen score, where finding $\pi$ such that $S(\pi,\theta)$ being a constant reduces to solving a second-order differential equation.

In model selection problems, since we consider loss functions $L:\calM\times \calM \to [0,\infty)$ where the outcome space and the action space are the same as $\calM$, the risk function $R^\pi(\hat{M})$ itself can be thought of as a scoring rule $S(\pi,\hat{M})$ that quantifies the distribution $\pi$ when we have an estimate $\hat{M}$. The following proposition describes the properness of the risk function as a scoring rule, relative to the family of point masses.

\begin{proposition}
\label{prop:scoringrule}
Let $L:\calM\times \calM \to [0,\infty)$ be a loss function that satisfies $L(M, \hat{M}) = 0$ if and only if $M = \hat{M}$ (identity of indiscernibles). Then the risk function $S(\pi, \cdot) = R^\pi(\cdot)$ is a strictly proper scoring rule relative to the family of point masses $\calF_\delta = \{\pi: \pi(M) = 1 \text{ at some } M\in \calM \}$.
\end{proposition}

Although the family of point mass distributions $\calF_\delta$ is a very narrow subset among all possible distributions on $\calM$, the risk function as a scoring rule being strictly proper relative to $\calF_\delta$ provides a natural interpretation in Bayesian model selection problems. 
Unlike Bayesian model averaging \citep{Hoeting1999-ew}, the model selection problem typically assumes, either explicitly or implicitly, that there is a true model that generated data. 
For example, \citet{Bayarri2012-ho} outlines criteria to be satisfied by objective priors for model selection problems, including model selection consistency and information consistency, describing the behavior of posterior probabilities and Bayes factors given the true data-generating model. 
Thus, it is natural to consider the family of point mass distributions $\calF_\delta$ if we believe that the true model belongs to $\calF_\delta$.
The risk function being strictly proper relative to $\calF_\delta$ implies that the Bayes action, which quotes a single model that minimizes the risk, is the unique optimal action if data are indeed generated from that quoted model. The only requirement of \cref{prop:scoringrule} is that the loss function satisfies the identity of indiscernibles and positive otherwise.

Depending on the choice of loss $L$ or risk function $R$, the risk equilibrium prior may not exist, or does not have full support (i.e. assigning zero prior probabilities to some models). We collect such counterexamples in Appendix~\ref{appendix:counterexamples}, which happens when the loss function $L$ is deliberately chosen so that it does not ``conform'' to the geometry of the model space.

\subsection{Examples of risk equilibrium priors in BVS}
Consider BVS problem with model space $\calM = \{0,1\}^p$. Under the zero-one loss $L_{01}$, it is straightforward to see that the only possible risk equilibrium prior with respect to $L_{01}$ is the uniform prior $p(\bm\gamma) = 2^{-p}$. This fact is not only limited to the BVS but also applies to all model selection problems with finite model space, even for unstructured cases, where uniform prior is the unique risk equilibrium prior with respect to the zero-one loss.
Next, the following theorem characterizes the risk equilibrium prior with respect to the generalized Hamming loss, providing necessary and sufficient conditions to become a risk equilibrium prior.

\begin{theorem}
\label{thm:ghloss_REprior}
A prior $\pi$ on the BVS model space $\calM = \{0,1\}^p$ is a risk equilibrium prior with respect to generalized Hamming loss $L_{\mathrm{GH}(a)}$ if and only if $\bbP^\pi(\gamma_i = 1) = 0.5a$ for $i=1,\dots,p$. 
\end{theorem}

\begin{example}[Uniform and multiplicity-adjusted priors]
The risk equilibrium prior with respect to $L_{\mathrm{H}}$ surpasses many existing popular objective priors that have been proposed in BVS literature. It only requires the prior marginal inclusion probabilities to be 0.5 for all variables. An obvious example is the uniform prior $p(\bm\gamma) = 2^{-p}$ by the symmetry of $\calM$. Another example is the beta-binomial prior that arises from a fully Bayesian treatment of a prior inclusion probability \citep{Cui2008-iu,Ley2009-ur}, also called multiplicity-adjusted prior \citep{Scott2010-rp},
\begin{equation}
\label{eq:betabin}
    \gamma_i\given \omega\iidsim \mathrm{Bernoulli}(\omega),\quad  i=1,\dots,p, \qquad \omega\sim \mathrm{Beta}(a_\omega,b_\omega).
\end{equation}
When $a_\omega = b_\omega$, the marginal inclusion probabilities are all 0.5, which becomes a risk equilibrium prior with respect to $L_{\mathrm{H}}$. It assigns uniform prior probabilities to model sizes when $a_\omega = b_\omega = 1$.
\end{example} 

\begin{example}[Asymmetric beta-binomial priors] For more general cases with unequal beta shape parameters, since prior marginal inclusion probabilities are $\bbP(\gamma_i=1)= \bbE[\omega]=a_\omega/(a_\omega + b_\omega)$ for all $i=1,\dots,p$, the beta-binomial prior with shape parameters satisfying $a_\omega/(a_\omega + b_\omega) = 0.5a$, such as $(a_\omega,b_\omega) = (a,2-a)$, is a risk equilibrium prior with respect to GH(a) loss.    
\end{example}

Although uniform and symmetric beta-binomial are both risk equilibrium priors with respect to $L_\mathrm{H}$, it does not imply that two priors have the same property in terms of model selection results with $L_{\mathrm{H}}$ \citep{Scott2010-rp}, and we note that prior risk behavior simply serves as a role of summary statistics through the lens of loss function. Nevertheless, if we consider the same subclass of loss functions and priors such as $L_{\mathrm{GH}(a)}$ and beta-binomial priors, risk equilibrium priors can serve as calibration purposes; we defer the detailed discussion to Section 5 with simulation study.

\subsection{Examples of risk equilibrium priors in BCA}
\label{subsec:bcariskeq}
Next, consider BCA problem with model space $\calM = \calP_{[p]}$, a collection of all possible partitions of $\{1,\dots,p\}$. Before we proceed, we refine our interest of priors to \textit{exchangeable} random partition priors \citep{Pitman1995-zc}, so that $\pi$ is invariant of the permutations of data indices (it is different from the permutation of labels $z_i$, which corresponds to different representation of a same partition). Exchangeability is a standard prior condition in BCA, unless one knows a priori that the data have some dependency structure, e.g. indexed by time or came from heterogeneous sources. 
In the following theorem, we characterize risk equilibrium priors under the generalized Binder's loss.

\begin{theorem} 
\label{thm:gbloss_REprior}
An exchangeable prior $\pi$ on the BCA model space $\calM= \calP_{[p]}$ is a risk equilibrium prior with respect to generalized Binder's loss $L_{\mathrm{GB}(a)}$ if and only if $\pi$ has a prior co-clustering probability $\bbP^\pi(z_i=z_{j}) = (2-a)/2$ for all $(i,j)$, $i\neq j$.
\end{theorem}

We provide several examples of risk equilibrium priors with respect to generalized Binder's loss that induces different behaviors of the prior number of clusters. These correspond to a family of Gibbs-type priors \citep{De_Blasi2015-xt}, and are implicitly assumed in popular Bayesian mixture models including the Dirichlet process mixture models \citep{Lo1984-ej}.

\begin{example}[Chinese restaurant process]
\label{ex:crp}
The first example is the Chinese restaurant process (CRP) with parameter $\theta>0$, denoted as CRP($\theta$), an exchangeable random partition prior induced by the Dirichlet processes \citep{Ferguson1973-av}. It is well known that the number of clusters under CRP has a logarithmic growth in terms of $p$. The prior $\pi$ can be sequentially defined as follows: first let $z_1 = 1$, and if $\bfz_{1:i}$ has $k$ clusters with sizes $n_1,\dots,n_k$, allocate label $z_{i+1}$ with probabilities $\bbP(z_{i+1} = j \given \bfz_{1:i}) = n_j/(\theta +i)$ for $j=1,\dots,k$ and $\bbP(z_{i+1} = k+1 \given \bfz_{1:i}) = \theta/(\theta +i)$, $i=1,\dots,p-1$. The CRP with parameter $\theta = a/(2-a)$ is a risk equilibrium prior with respect to $L_{\mathrm{GB}(a)}$, since the prior co-clustering probability of CRP($\theta$) is $\bbP(z_2 = 1 \given z_1 = 1) = 1/(\theta+1)$ by exchangeability. 
One can also consider a generalized gamma prior on $\theta$  which comes with attractive properties \citep{Ascolani2022-nd}, as long as it satisfies $\bbE[1/(\theta+1)] = (2-a)/2$. Such an example is a Weibull distribution with shape parameter 2 and scale parameter approximately $1.3115$. 
Conjugate priors \citep{Zito2023-yq} could be also considered, if cluster inference is a main purpose but not a prediction.

\end{example}

\begin{example}[Two-parameter CRP] The second example is the two-parameter CRP with parameters $(\sigma,\theta)$ with $\sigma\in(0,1)$ and $\theta > -\sigma$,  denoted as CRP2($\sigma,\theta$), an exchangeable random partition induced by the Pitman-Yor processes \citep{Perman1992-yh,Pitman1997-ym}. Under CRP2, the number of clusters has a fractional power growth in terms of $p$. 
The prior can be defined as follows: first let $z_1 = 1$, and if $\bfz_{1:i}$ has $k$ clusters with sizes $n_1,\dots,n_k$, allocate label $z_{i+1}$ with probabilities $\bbP(z_{i+1} = j \given \bfz_{1:i}) = (n_j-\sigma)/(\theta +i)$ for $j=1,\dots,k$ and $\bbP(z_{i+1} = k+1 \given \bfz_{1:i}) = (\theta + k\sigma)/(\theta +i)$, $i=1,\dots,p-1$. 
The CRP2($\sigma, (a-2\sigma)/(2-a)$) is a risk equilibrium prior with respect to $L_{\mathrm{GB}(a)}$, since the prior co-clustering probability under CRP2($\sigma, \theta$) is $(1-\sigma)/(\theta+1)$. 
\end{example}

\begin{example}[Mixture of Dirichlet-multinomial]
\label{ex:dirimultimixture}
Another popular random partition prior is the mixture of Dirichlet-multinomial, which is induced from the mixture of finite mixture model \citep{Miller2018-ph} by introducing a prior on the number of components $K$. We emphasize that the number of components $K$ is different from the number of clusters, where the latter corresponds to the number of ``filled'' components. Unlike CRP, the induced distribution of the number of clusters remains finite as $p$ grows. The prior $\pi$ can be hierarchically defined as $z_i \given \bfp,K \iidsim \mathrm{Categorical}(\bfp)$, $\bfp \given K\sim \mathrm{Dirichlet}_K(\alpha,\dots,\alpha)$, and $K\sim q_K$ for some distribution $q_K$ supported on $\bbN$. When 
$K\sim q_K$ satisfies $\bbE^{q_K}[(1+\alpha)/(\alpha K + 1)] = (2-a)/2$, then $\pi$ is a risk equilibrium prior with respect to $L_{\mathrm{GB}(a)}$. 
For example when $a=1$ and $\alpha = 1$, $q_K$ being a shifted Poisson distribution (so that $K-1$ follows Poisson) with parameter $\lambda \approx 2.5569$ satisfies the criterion.
\end{example}

\begin{example}[Balance-neutral random partition]
\label{ex:balanceneutral}
The final example is the balance-neutral random partition priors \citep{Lee2022-yk}, which induce a uniform prior on partitions conditionally on the number of clusters so that probabilities only depend on the number of clusters and neutral to the balancedness of the cluster sizes. The prior $\pi$ is defined as $z_i\given K\iidsim \mathrm{Categorical}(K^{-1}, \dots,K^{-1})$ and $K\sim q_K$, which can be thought of as a limiting case of \cref{ex:dirimultimixture} when $\alpha \to \infty$ \citep{Gnedin2005-do}. When $K\sim q_K$ satisfies $\bbE[K^{-1}] = (2-a)/2$, which corresponds to the prior co-clustering probability, the resulting $\pi$ is a risk equilibrium prior with respect to $L_{\mathrm{GB}(a)}$. For example when $a=1$, $q_K$ being a geometric distribution with a success probability of 0.2847 gives $\bbE[K^{-1}]\approx 0.5$.
\end{example}

The elicitation of objective priors for Bayesian cluster analysis problem is rarely discussed in the literature, partly due to the fact that ``objectivity'' is not well defined and the model space $\calP_{[p]}$ of BCA has a much more complex structure than a hypercube of BVS. To the best of our knowledge, only \citet{Casella2014-tb} proposed hierarchical uniform prior as an objective choice for BCA, by first assigning the prior on the number of clusters (not components) and choosing a partition uniformly at random conditionally on the number of clusters, where the conditional uniform assignment is the main ground for objectivity. However, this construction is different from the balance-neutral prior described in \cref{ex:balanceneutral} based on the prior on the number of components $K$. A potential limitation of the hierarchical uniform prior is that there is no guarantee of projectivity, also known as the addition rule or Kolmogorov consistency, which is an essential property of random partition prior when it comes to prediction purposes. We refer \cite{Betancourt2022-qc} for a detailed discussion on the role of the projectivity assumption. Nonetheless, if cluster inference is a main purpose but not a prediction, the hierarchical uniform prior could be used, and it is a risk equilibrium prior with respect to $L_{\mathrm{B}}$ when it induces co-clustering probability 0.5. The balance-neutral prior described in \cref{ex:balanceneutral} characterizes a complete family of exchangeable prior that satisfies projectivity and has uniform probabilities given the number of clusters; see \citet{Lee2022-yk} for details.

While those examples all belong to risk equilibrium priors with respect to generalized Binder's loss, each induces different asymptotic behavior on the number of clusters and can be chosen based on the different application scenarios. For example, when the number of clusters is expected to grow logarithmically, CRP prior in \cref{ex:crp} can be chosen as an objective prior with respect to Binder's loss that reflects prior belief on the logarithmic growth of the number of clusters. When the true number of clusters is assumed to be finite, \cref{ex:dirimultimixture} can be used, and to further align with the notion of objectivity of \cite{Casella2014-tb} based on conditional uniform assignment, we recommend balance-neutral priors in \cref{ex:balanceneutral} which come with projectivity guarantee.

It is worth noting that all risk equilibrium priors in previous examples do not put equal prior probabilities on the number of clusters (except hierarchical uniform), but typically induce an unimodal distribution on the number of clusters. For example under CRP(1), the distribution of the number of clusters becomes similar to Poisson with parameter $(\log p - \psi(1))$ shifted by 1 as $p$ grows \citep{West1992-zo}, where $-\psi(1)\approx 0.5772$ is an Euler's constant. This looks like a penalizing prior by having prior probabilities on the number of clusters mainly around $\log p \ll p$, but the prior risk with respect to $L_{\mathrm{B}}$ is a constant function, and risk equilibrium prior provides a natural adjustment that takes into account the geometry of model space $\calP_{[p]}$ through $L_{\mathrm{B}}$. 
We conclude this section with the following counterexample, illustrating the uniform prior on a partition space is not a risk equilibrium prior with respect to symmetric Binder's loss $L_{\mathrm{B}}$.

\begin{remark}[Uniform is not risk equilibrium with respect to $L_{\mathrm{B}}$]
We note that the uniform distribution over the partition space, $\pi(\bm\rho) = 1/\scrB_p$ with  $\scrB_p$ be a $p$th Bell number, is not a risk equilibrium prior for any $p$ with respect to symmetric Binder's loss $L_{\mathrm{B}}$. The case $p=3$ gives a counterexample, where $R^\pi_B(\{\{1,2,3\}\}) = 9/5$, $R^\pi_B(\{\{1\},\{2,3\}\}) = R^\pi_B(\{\{2\},\{1,3\}\}) = R^\pi_B(\{\{3\},\{1,2\}\})= 7/5$, and $R^\pi_B(\{\{1\},\{2\},\{3\}\}) = 6/5$. The prior risk is minimized at the most complex model, the partition of singletons, which is an undesirable prior risk behavior associated with $L_{\mathrm{B}}$.
\end{remark}

\section{Risk penalization priors with respect to a loss}

\subsection{Definition and properties}
Similar to the risk equilibrium priors, we introduce the risk penalization priors where the prior risk becomes an increasing function with respect to the partial order $\prec$ that compares model complexity. 

\begin{definition}[Risk penalization prior] For a structured model space $\calM$ equipped with a partial order $\prec$, we say a prior $\pi$ is a \textit{risk penalization prior with respect to loss $L$, or with respect to risk $R^\pi$} when the prior risk $R^\pi(\cdot) = \bbE^{\pi}[L(M,\cdot)]$ satisfies $\hat{M}_1 \prec \hat{M}_2 \implies R^\pi(\hat{M}_1) \le R^\pi(\hat{M}_2)$.
\end{definition}

Risk penalization priors assign higher prior risk to models that are more complex, and the Bayes action selects the simplest model before seeing data (if unique). Similarly to risk equilibrium priors, risk penalization priors do not uniquely determine a probability distribution, but based on a given loss $L$ or risk $R$, they define a family of priors that have increasing prior risk properties in terms of model complexity. Here, note that the increasing prior risk in terms of $\prec$ is a sufficient but not a necessary condition for Bayes action to select the simplest model before observing the data. That is, Bayes action may still choose the simplest model even if the prior risk may decrease at some comparable pairs of models. Nevertheless, it is still natural to define risk penalization priors based on increasing prior risk in all comparable pairs since penalization of model complexity should be applied universally.

The family of risk penalization priors includes several popular penalization (sparsity-inducing) priors in Bayesian model selection problems, especially those tailored for high-dimensional model selection problems where regularization is necessary to achieve a stable fit and better predictive accuracy. 
Although marginal likelihoods $f(D\given M) = \int f(D\given \vartheta,M)p(\vartheta\given M) d\vartheta$ naturally penalize more complex models through Occam's razor effect, additional penalization for prior model probabilities are often employed to establish theoretical guarantees such as model selection consistency; see \citet{Narisetty2014-dg} in the context of BVS.

The notion of risk penalization priors and penalization on prior probabilities based on model complexity are not equivalent. Depending on the choice of loss function, decreasing prior probabilities does not necessarily imply an increasing prior risk in terms of $\prec$, and it is even possible to have a risk penalization prior with increasing prior probabilities in terms of $\prec$; see Appendix~\ref{appendix:counterexamples} for counterexamples.
One important exception is the risk penalization prior with respect to zero-one loss $L_{01}$, where risk penalization prior becomes equivalent to $\hat{M}_1 \prec \hat{M}_2 \implies \pi(\hat{M}_1) \ge \pi(\hat{M}_2)$, putting less prior probability to more complex models.

\subsection{Examples of risk penalization priors in BVS}
The following theorem characterizes the risk penalization priors with respect to generalized Hamming loss, which only depends on marginal inclusion probabilities.

\begin{theorem}
\label{thm:ghloss_RPprior}
A prior $\pi$ on the BVS model space $\calM = \{0,1\}^p$ is a risk penalization prior with respect to generalized Hamming loss $L_{\mathrm{GH}(a)}$ if and only if $\bbP^\pi(\gamma_i = 1) \le 0.5a$ for $i=1,\dots,p$. 
\end{theorem}

Combined with \cref{thm:ghloss_REprior}, it highlights that the notion of risk equilibrium and penalization prior is a relative concept that depends on the choice of the loss function.  
In other words, when prior $\pi$ is a risk penalization prior with respect to some loss, the same prior can be also interpreted as a risk equilibrium prior with respect to another loss. For example, the uniform prior $\pi(\bm\gamma) = 2^{-p}$ is a risk penalization prior with respect to generalized Hamming loss $L_{\mathrm{GH}(a)}$ with $a\in (1,2)$, but at the same time, it is also a risk equilibrium prior with respect to symmetric Hamming loss $L_{\mathrm{H}}$. 

\begin{example}[Beta-binomial priors and its variant]
\cite{Castillo2015-pl} considered beta-binomial model \eqref{eq:betabin} with parameters $a_\omega=1$ and $b_\omega=p^u$ with some $u>1$. This leads to the prior marginal inclusion probabilities $1/(p^u + 1)$, which is always less than 0.5. 
Also, \cite{Villa2020-bv} considered $\gamma_i\given \omega \iidsim \mathrm{Ber}(\omega)$, reparameterizing $\omega = 1/(1+e^\kappa)\in(0,0.5)$ for some $\kappa>0$ and considered prior on $\kappa$, which always gives marginal inclusion probability less than $0.5$. Thus, those priors based on Bernoulli trials are all risk penalization prior with respect to Hamming loss $L_{\mathrm{H}}$. Further, depending on the hyperparameter choice such as $u=2$ in the first example, it can also become a risk penalization prior with respect to generalized Hamming loss with some $a>1$.

\end{example}

\begin{example}[Exponentially decaying and truncated prior]
    \cite{Yang2016-yu} considered prior model probabilities that have an exponential decay and an upper bound $s_{\mathrm{max}}$ in model sizes, 
\begin{equation}
\label{eq:yangprior}
\pi(\bm\gamma) \propto p^{-\kappa |\bm\gamma|}\ind( |\bm\gamma|\le s_{\mathrm{max}})    
\end{equation}
for some $\kappa\ge 2$. The truncation of the model space is to establish rapid convergence results of a Markov chain Monte Carlo (MCMC) sampler; see also \cite{Chang2022-vo}. It can be shown that prior \eqref{eq:yangprior} is also a risk penalization prior with respect to $L_{\mathrm{H}}$, see Appendix \ref{appendix:yangRPproof} for details. 
\end{example}

\subsection{Examples of risk penalization priors in BCA}

We also provide examples of risk penalization priors in BCA. First, when prior $\pi$ is exchangeable, we characterize the risk penalization priors with respect to generalized Binder's loss, which only depends on prior co-clustering probability.

\begin{theorem}
\label{thm:gbloss_RPprior}
An exchangeable prior $\pi$ on the BCA model space $\calM = \calP_{[p]}$ is a risk penalization prior with respect to generalized Binder's loss $L_{\mathrm{GB}(a)}$ if and only if $\pi$ has a prior co-clustering probability $\bbP^\pi(z_i=z_{j}) \ge (2-a)/2$ for all $(i,j)$, $i\neq j$.
\end{theorem}

The examples of risk penalization priors with respect to generalized Binder's loss can be easily obtained from previous examples presented in Section~\ref{subsec:bcariskeq} by simply changing equality to inequality that leads to fewer number of clusters. For example, CRP$(\theta)$ prior with $0<\theta \le 1$ and CRP2($\sigma, \theta$) with $-\sigma < \theta \le (1-2\sigma)$ are risk penalization priors with respect to symmetric Binder's loss $L_{\mathrm{B}}$. 

The use of VI loss for Bayesian cluster analysis has become increasingly popular in recent years. 
\citet{Wade2018-mp} discussed the similarities and differences between Binder's loss and the VI loss, illustrating the Bayes estimator under Binder's loss typically suffers from an over-clustering problem compared to the VI loss. 
They explained this behavior by comparing loss functions at two extremes, partition of singletons and one-block partition. 
For example, Property 6 of \citet{Wade2018-mp} describes that a two-block partition with equal sizes is closer to a one-block partition under the VI loss, 
but at the same time it is closer to the partition of singletons under Binder's loss. This illustration can help us understand how Binder's loss and VI loss differently measure the similarity between two partitions, but the connection to risk functions and corresponding Bayes estimators is less clear. To better explain the phenomenon that the Bayes estimator under Binder's loss typically yields more clusters than VI loss, we compare prior risk functions' properties under the same prior but with different loss functions. 

In the previous example, the prior risk with respect to Binder's loss $R_B^\pi$ only depends on the prior co-clustering probability $c$ and the risk becomes constant when $c=0.5$. However, the prior risk with respect to VI loss $R^\pi_V(\hat{\bfz})$ in equation \eqref{eq:virisk} cannot be expressed as a function of $c$, but rather it involves expectation over nonlinear terms that involve prior $\pi$ and partition estimate $\hat{\bfz}$, leading to complicated prior risk analysis.  
Instead, its lower bound $\tilde{R}^\pi_V(\hat{\bfz})$ in equation \eqref{eq:virisklb} only depends on the prior co-clustering probability $c$, which greatly simplifies prior risk analysis and allows to make a risk comparison within a family of prior distribution with same $c$. In the following, we show that if and only if $c\ge \sqrt{2}-1$, an exchangeable prior $\pi$ is a risk penalization prior with respect to the VI risk lower bound $\tilde{R}^\pi_V$, in other words, $\tilde{R}^\pi_V(\hat\bfz)$ is an increasing function of $\hat\bfz$ in terms of $\prec$.

\begin{figure}
    \centering
    \includegraphics[width=0.8\textwidth]{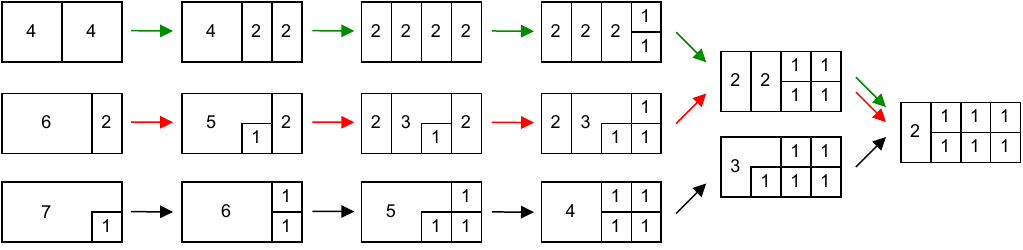}
    \includegraphics[width=0.9\textwidth]{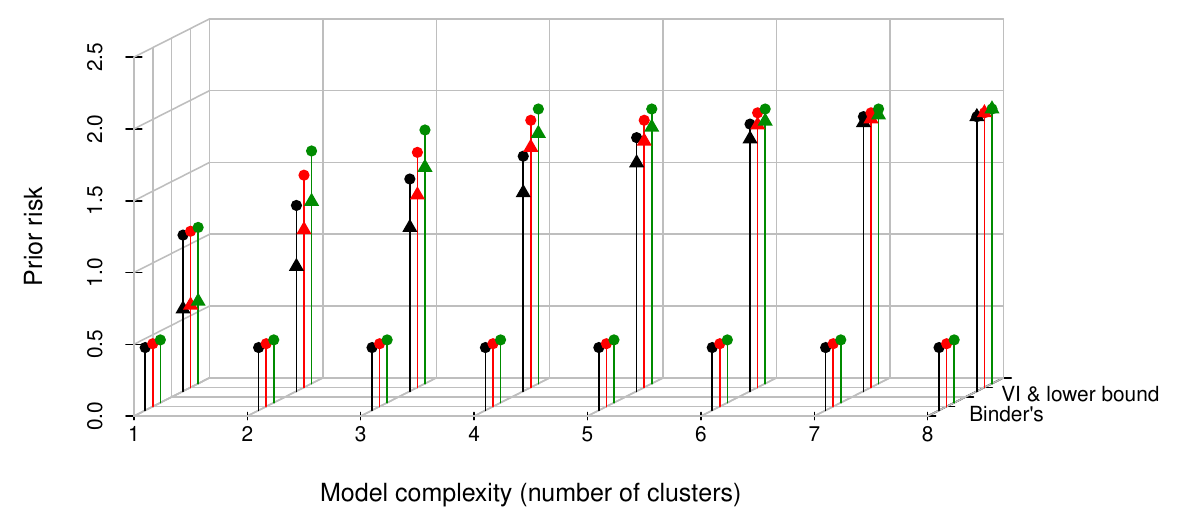}
    \caption{Illustration of risk equilibrium and penalization priors in BCA. (Top) Each rectangle box represents a partition of $\{1,\dots,8\}$ with cluster sizes shown in numbers. The arrangement and green, red, black arrows represent three chains of partitions, which are ordered with respect to $\prec$. One-block partition at the beginning and partition of singletons at the end are omitted. (Bottom) Under the same CRP(1) prior $\pi$ on the whole partition space $\calP_{[8]}$, two different prior risks are evaluated at three chains: (Bottom-front) Prior risk with respect to Binder's loss $R^\pi_B$ which is a constant (divided by $32$ for visualization). (Bottom-back) Prior risk with respect to VI loss $R^\pi_V$ shown in circles and its lower bound $\tilde{R}^\pi_V$ shown in triangles, both increase with respect to $\prec$.}
    \label{fig:riskplot}
\end{figure}

\begin{theorem}
\label{thm:viloss_RPprior}
An exchangeable prior $\pi$ on the BCA model space $\calM= \calP_{[p]}$ is a risk penalization prior with respect to $\tilde{R}_{V}$ if and only if $\pi$ has a prior co-clustering probability $\bbP^\pi(z_i=z_{j}) \ge \sqrt{2}-1 \approx 0.414$ for all $(i,j)$, $i\neq j$.
\end{theorem}

Under the same CRP(1) prior which induces prior co-clustering probability $c=0.5$, the prior risks $R_B^\pi$ (Binder's), $R_V^\pi$ and its lower bound $\tilde{R}_V^\pi$ are visualized in Figure~\ref{fig:riskplot}. As the number of clusters grows, the prior risk with respect to VI loss and its lower bound increases, suggesting that $R_V^\pi$ and its lower bound penalize more complex models before seeing data, which clearly shows a different pattern from $R_B^\pi$ which is a constant function. The interpretation based on prior risks is consistent with the findings of \citet{Wade2018-mp}, where both the VI risk and its lower bound minimizer can mitigate the over-clustering problem of Binder's risk minimizer.

\section{Data analysis}

\subsection{Simulation studies}

We perform simulation studies in BVS settings similar to \citet{Scott2010-rp}, to support the role of risk equilibrium priors as objective priors with respect to a given loss in small $n$ settings. We set a small number of variables $p = 14$, where the corresponding size of the model space is $|\calM| = 2^{14} = 16834$. This size is large enough to be interesting, but also allows exact calculation of posterior model probabilities by enumeration, to avoid potential confounding effects due to computational issues such as a mixing problem of the MCMC algorithm. Under the normal linear model $\bm{y} = \bfX\bm\beta + \bm\epsilon$, we set true coefficients be $\bm\beta = (1,1,1,-1,-1,-1,\bm{0}_{8})^\top\in\bbR^{14}$ so that the number of nonzero variables is 6. The elements of design matrix $\bfX \in \bbR^{n\times p}$ are generated from i.i.d. standard normal and noise variance is set as $9$.  With three different $n\in \{20,60,1000\}$ representing small, moderate, and high sample size scenarios, we generated $1000$ replicated datasets. 

We consider total 8 loss functions, generalized Hamming loss $L_{\mathrm{GH}(a)}$ with 7 different weights $a\in \{0.7, 0.8, \dots, 1.3\}$ and zero-one loss $L_{01}$. As shown in \cref{thm:ghloss_REprior}, the corresponding risk equilibrium priors are those with marginal inclusion probability $0.5a$ for $L_{\mathrm{GH}(a)}$ and uniform prior $\pi_{\mathrm{unif}}$ for $L_{01}$, and for the former we choose beta-binomial priors \eqref{eq:betabin} with hyperparameter $(a_\omega,b_\omega) = (a,2-a)$, denoted as $\pi_{0.7},\dots, \pi_{1.3}$. We considered Jeffreys-Zellner-Siow prior for model parameters, the Jeffreys prior on intercept and noise variance and put Zellner-Siow prior \citep{Liang2008-ue} on $\bm\beta$ conditionally on the intercept and noise variance. Based on those prior settings, the posterior model probabilities are calculated by enumeration and the model selection is performed with Bayes estimators based on 8 different losses using \cref{prop:ghloss_riskeq} and posterior mode, resulting in a total of 64 configurations. All computations are carried out with the R package \texttt{BAS} \citep{Clyde2022-uj}.

To see how model selection results differ by the choice of prior and loss, we summarized simulation results in \cref{table:bvssim} based on the Hamming distance $L_{\mathrm{H}}$ from the reference Bayes estimator, which is based on prior $\pi_{1.0}$ and symmetric Hamming loss $L_{\mathrm{H}}$ (median probability model with $\pi_{1.0}$). 
The result shows different patterns based on a sample size $n$ which we elaborate on. When $n$ is small, the risk equilibrium priors with respect to $L_{\mathrm{GH}(a)}$ provide model selection results that are closest to the reference Bayes estimator for each loss function. 
In other words, choosing risk equilibrium priors within the same subclass provides the model selection results that are less sensitive to the choice of loss, supporting our main research question Q1 in Figure~\ref{fig:keyq}(a). As the sample size $n$ becomes moderate, the prior influence on the posterior model probabilities diminishes, and the model selection results are more heavily dependent on the choice of loss than prior. This can be seen from the results that, given a loss function, the smallest Hamming distance from the reference estimator is achieved at priors that encode stronger information than risk equilibrium priors. We also refer to the rows that contain the reference estimator under the same $L_{\mathrm{GH}(1.0)}$ loss, where the Hamming distance from the reference estimator quickly reduces as $n$ increases, showing the diminishing prior influence on the Bayes estimator which is desired.  
When $n$ is large enough so that posterior model probabilities almost concentrate on the true model, the model selection results almost coincide in all configurations regardless of the choice of prior and loss, as can be seen from the fact that the average Hamming distance from the reference model is less than $0.3$ in all settings. 

The results with the zero-one loss $L_{01}$ and uniform prior $\pi_{\mathrm{unif}}$, which are different subclasses of loss and priors from generalized Hamming loss and beta-binomial priors, show a completely different pattern. When $n$ is small, model selection results with the highest probability model (based on $L_{01}$) are not similar to the results obtained from the median or quantile probability models, which is expected behavior since the posterior distribution may exhibit multimodality due to the small sample size. 
However when $n$ becomes large, the posterior mode becomes closer to the results from median or quantile probability models, and interestingly the posterior mode is closest to the reference estimator under the uniform prior, the risk equilibrium prior with respect to $L_{01}$. In addition, although $\pi_{1.0}$ and $\pi_{\mathrm{unif}}$ are both risk equilibrium priors with respect to  $L_{\mathrm{GH}(1)}$, the model selection results differ in small to moderate $n$, which are consistent with the result of \citet{Scott2010-rp}. 
We emphasize that the constant prior risk property simply serves as a role of summary statistics of prior through the loss function, which is useful for calibration within the same subclass of loss and prior, but for other cases there is no guarantee of the Bayes estimator being similar since the landscape of posterior risk might be completely different under the different subclass of priors or loss. 
We present an additional table that summarizes the number of selected variables in Appendix~\ref{appendix:moretable}, where it can be seen that risk penalization priors always yield more parsimonious results than risk equilibrium priors.

\begin{table}[htbp]
\small
    \centering
   \caption{Hamming distances from the reference Bayes estimator (with loss $L_{\mathrm{GH}(1)}$ and prior $\pi_{1.0}$), averaged over 1000 replicates in BVS settings with three different $n$. $L_{\mathrm{GH}(a)}$ denotes generalized Hamming loss with weight $a$ and $L_{01}$ denotes zero-one loss. Underlines indicate results from risk equilibrium priors, and bolds indicate results that give the closest model to the reference model for a given loss.}
   \label{table:bvssim}
    \begin{tabular}{c cccccccc c}
    \toprule
            &  \multirow{2}{*}{Loss}  & \multicolumn{8}{c}{Model priors ($\pi_a \equiv$  Beta-Binomial$(a, 2-a)$)}  \\
      &    & $\pi_{0.7}$ & $\pi_{0.8}$ & $\pi_{0.9}$ & $\pi_{1.0}$ &$\pi_{1.1}$ & $\pi_{1.2}$ & $\pi_{1.3}$ & $\pi_{\mathrm{unif}}$\\
        \midrule
    \multirow{8}{*}{$n=20$}  & $L_{\mathrm{GH}(0.7)}$ & \underline{\bf 0.543} & 0.696 & 1.285 & 2.011 & 2.801 & 3.643 & 4.602 & 6.212 \\ 
      &  $L_{\mathrm{GH}(0.8)}$ &0.579 & \underline{\bf 0.334} & 0.596 & 1.208 & 1.940 & 2.703 & 3.582 & 4.531 \\ 
      & $L_{\mathrm{GH}(0.9)}$ & 0.901 & 0.475 & \underline{\bf 0.175} & 0.533 & 1.187 & 1.917 & 2.689 & 3.492 \\ 
      & $L_{\mathrm{GH}(1.0)}$ &  1.238 & 0.861 & 0.470 & \underline{\bf Ref.} & 0.509 & 1.137 & 1.921 & \underline{2.837} \\ 
      &  $L_{\mathrm{GH}(1.1)}$ & 1.497 & 1.214 & 0.844 & 0.486 & \underline{\bf 0.155} & 0.535 & 1.169 & 2.324 \\ 
      & $L_{\mathrm{GH}(1.2)}$ & 1.699 & 1.478 & 1.196 & 0.875 & 0.498 & \underline{\bf 0.340} & 0.622 & 1.986 \\ 
      &  $L_{\mathrm{GH}(1.3)}$ &  1.910 & 1.703 & 1.469 & 1.209 & 0.903 & 0.584 & \underline{\bf 0.514} & 1.755 \\ 
      \cmidrule{2-10}
      &  $L_{01}$ & 2.182 & 1.862 & 1.607 & \bf 1.477 & 1.590 & 1.939 & 2.524 & \underline{2.837} \\ 
        \midrule
    \multirow{7}{*}{$n=60$}  & $L_{\mathrm{GH}(0.7)}$ & \underline{\bf 0.714} & 1.116 & 1.536 & 2.005 & 2.561 & 3.174 & 3.874 & 2.713 \\ 
      & $L_{\mathrm{GH}(0.8)}$ &  \bf 0.262 & \underline{0.426} & 0.784 & 1.205 & 1.692 & 2.154 & 2.810 & 2.252 \\
      & $L_{\mathrm{GH}(0.9)}$&  0.422 & \bf 0.158 & \underline{0.204} & 0.550 & 0.913 & 1.403 & 1.927 & 1.897 \\  
      & $L_{\mathrm{GH}(1.0)}$ & 0.873 & 0.602 & 0.296 & \underline{\bf Ref.} & 0.331 & 0.715 & 1.193 & \underline{1.687} \\ 
      & $L_{\mathrm{GH}(1.1)}$ & 1.238 & 1.014 & 0.755 & 0.488 & \underline{0.198} & \bf 0.186 & 0.531 & 1.593 \\ 
      & $L_{\mathrm{GH}(1.2)}$ &  1.567 & 1.375 & 1.154 & 0.925 & 0.666 & \underline{0.387} & \bf 0.238 & 1.576 \\ 
      & $L_{\mathrm{GH}(1.3)}$& 1.880 & 1.683 & 1.496 & 1.304 & 1.081 & 0.828 & \underline{\bf 0.574} & 1.563 \\ 
      \cmidrule{2-10}
      &  $L_{01}$ & 2.274 & 2.209 & 2.197 & 2.250 & 2.484 & 2.941 & 3.569 & \underline{\bf 1.664} \\ 
        \midrule
          \multirow{7}{*}{$n=1000$}  & $L_{\mathrm{GH}(0.7)}$ & \underline{\bf 0.163} & 0.173 & 0.188 & 0.208 & 0.221 & 0.251 & 0.277 & 0.147 \\ 
      &  $L_{\mathrm{GH}(0.8)}$ &  \bf 0.092 & \underline{0.105} & 0.113 & 0.125 & 0.139 & 0.155 & 0.175 & 0.073 \\ 
      &  $L_{\mathrm{GH}(0.9)}$ &  \bf 0.015 & 0.031 & \underline{0.041} & 0.055 & 0.066 & 0.085 & 0.097 & 0.031 \\ 
      &  $L_{\mathrm{GH}(1.0)}$ &0.028 & 0.021 & 0.008 & \underline{\bf Ref.} & 0.006 & 0.014 & 0.028 & \underline{0.045} \\ 
      &  $L_{\mathrm{GH}(1.1)}$ &0.071 & 0.063 & 0.057 & 0.049 & \underline{0.040} & 0.032 & \bf 0.017 & 0.072 \\ 
      &  $L_{\mathrm{GH}(1.2)}$ & 0.094 & 0.089 & 0.086 & 0.085 & 0.078 & \underline{0.071} & \bf 0.066 & 0.097 \\ 
      &  $L_{\mathrm{GH}(1.3)}$ & 0.112 & 0.108 & 0.105 & 0.105 & 0.101 & 0.094 & \underline{\bf 0.089} & 0.115 \\ 
        \cmidrule{2-10}
      &  $L_{01}$ & 0.081 & 0.077 & 0.072 & 0.068 & 0.064 & 0.059 & 0.054 & \underline{\bf 0.051} \\ 
      \bottomrule
    \end{tabular}
\end{table}

\subsection{Real data analysis}

To see how risk equilibrium priors and risk penalization priors affect the Bayes estimator in BCA, we consider galaxy data, the dataset of velocities of 82 distant galaxies (in 1000km/s) that is publicly available in the R package \texttt{MASS} \citep{Venables2002-ow}. This dataset serves as a popular benchmark data in Bayesian cluster analysis, see \citet{Grun2022-dq} for previous clustering studies that used the galaxy data. 

Similar to the simulation studies, we consider generalized Binder's loss $L_{\mathrm{GB}(a)}$ with 7 different weights $a\in \{0.7, 0.8, \dots,1.3\}$. As shown in \cref{thm:gbloss_REprior}, the corresponding risk equilibrium priors for GB(a) loss are the ones with prior co-clustering probability $c = (2-a)/2$, and we choose CRP($\theta$) prior with different hyperparameter choices $\theta =  a/(2-a)$ denoted as $\pi_{0.7},\dots, \pi_{1.3}$, which is equivalent to fitting a DP mixture model with concentration parameter $\theta =  a/(2-a)$. We consider the mixture of normals, with unknown location and scale parameters and further placed hyperpriors based on default settings of the R package \texttt{BNPmix} \citep{Corradin2021-ag}. Using the \texttt{BNPmix} R package, we ran the posterior inference algorithm \citep{Canale2022-fp} with 100,000 MCMC iterations and discarded the first 20,000 samples as burn-in. 
With 80,000 posterior samples, we obtained Bayes estimators using a randomized greedy search algorithm with the R package \texttt{salso} \citep{Dahl2022-qd}. In addition, we also calculated Bayes estimators based on VI risk $R^\Pi_V$ and its lower bound $\tilde{R}^\Pi_V$ so that total $9\times 7 = 63$ Bayes estimators are collected, and the whole process is repeated 10 times on the same dataset.

\begin{table}
\small
    \centering
   \caption{Binder's distances from the reference Bayes estimator (with loss $L_{\mathrm{GB}(1)}$ and prior $\pi_{1.0}$), averaged over 10 repeated fits on the galaxy data. $L_{\mathrm{GB}(a)}$ denotes generalized Binder's loss with weight $a$. Underlines indicate results from risk equilibrium priors, and bolds indicate results that give the closest model to the reference model for a given loss.}
   \label{table:galaxydist}
    \begin{tabular}{cccccccc}
    \toprule
         Loss  & \multicolumn{7}{c}{Model priors, $\pi_a \equiv$  CRP$(a/(2-a))$} \\
         (or risk) & $\pi_{0.7}$ & $\pi_{0.8}$ & $\pi_{0.9}$ & $\pi_{1.0}$ &$\pi_{1.1}$ & $\pi_{1.2}$ & $\pi_{1.3}$ \\
        \midrule
$L_{\mathrm{GB}(0.7)}$ & \underline{\bf 1.0} & 63.1 & 138.0 & 138.0 & 205.0 & 205.0 & 342.4 \\ 
$L_{\mathrm{GB}(0.8)}$ & \bf 0.0 & \underline{1.0} & 1.0 & 138.0 & 138.0 & 191.6 & 205.0 \\ 
$L_{\mathrm{GB}(0.9)}$ & 56.4 & \bf 0.0 & \underline{\bf 0.0} & 1.0 & 117.6 & 138.0 & 198.3 \\ 
$L_{\mathrm{GB}(1.0)}$& 140.0 & 140.0 & 56.6 & \underline{\bf Ref.} & 0.0 & 70.0 & 138.0 \\ 
$L_{\mathrm{GB}(1.1)}$& 140.0 & 140.0 & 140.0 & 140.0 & \underline{\bf 0.0} & \bf 0.0 & 97.2 \\ 
$L_{\mathrm{GB}(1.2)}$& 140.0 & 140.0 & 140.0 & 140.0 & 140.0 & \underline{14.0} & \bf 0.0 \\ 
$L_{\mathrm{GB}(1.3)}$& 140.0 & 140.0 & 140.0 & 140.0 & 140.0 & 140.0 & 
\underline{\bf 112.2} \\ 
\midrule
$R^\Pi_{V}$& 140.0 & 140.0 & 140.0 & 140.0 & 140.0 & 140.0 & 140.0 \\ 
$\tilde{R}^\Pi_{V}$& 140.0 & 140.0 & 140.0 & 140.0 & 140.0 & 140.0 & 140.0 \\ 
      \bottomrule
    \end{tabular}
\end{table}

\begin{table}
\small
    \centering
   \caption{The number of clusters in Bayes estimators, averaged over 10 repeated fits on the galaxy data. Underlines indicate results from risk equilibrium priors, and the lower diagonal part (excluding last two rows) corresponds to risk penalization priors.}
   \label{table:galaxyclust}
    \begin{tabular}{cccccccc}
    \toprule
         Loss  & \multicolumn{7}{c}{Model priors, $\pi_a \equiv$  CRP$(a/(2-a))$} \\
         (or risk) & $\pi_{0.7}$ & $\pi_{0.8}$ & $\pi_{0.9}$ & $\pi_{1.0}$ &$\pi_{1.1}$ & $\pi_{1.2}$ & $\pi_{1.3}$ \\
        \midrule
$L_{\mathrm{GB}(0.7)}$ & \underline{5.0} & 5.9 & 7.0 & 7.0 & 8.0 & 8.0 & 10.1 \\ 
$L_{\mathrm{GB}(0.8)}$ & 4.0 & \underline{5.0} & 5.0 & 7.0 & 7.0 & 7.8 & 8.0 \\
$L_{\mathrm{GB}(0.9)}$ & 3.8 & 4.0 & \underline{4.0} & 5.0 & 6.7 & 7.0 & 7.9 \\ 
$L_{\mathrm{GB}(1.0)}$& 3.0 & 3.0 & 3.9 & \underline{4.0} & 4.0 & 6.0 & 7.0 \\ 
$L_{\mathrm{GB}(1.1)}$& 3.0 & 3.0 & 3.0 & 3.0 & \underline{4.0} & 4.0 & 6.4 \\ 
$L_{\mathrm{GB}(1.2)}$& 3.0 & 3.0 & 3.0 & 3.0 & 3.0 & \underline{3.9} & 4.0 \\ 
$L_{\mathrm{GB}(1.3)}$&  3.0 & 3.0 & 3.0 & 3.0 & 3.0 & 3.0 & \underline{3.3} \\ 
\midrule
$R^\Pi_{V}$& 3.0 & 3.0 & 3.0 & 3.0 & 3.0 & 3.0 & 3.0 \\ 
$\tilde{R}^\Pi_{V}$& 3.0 & 3.0 & 3.0 & 3.0 & 3.0 & 3.0 & 3.0 \\ 
      \bottomrule
    \end{tabular}
\end{table}

\cref{table:galaxydist} shows how clustering results with the Bayes estimator differ by the choice of prior and loss functions. Here the difference of clustering is measured with Binder's distance $L_{\mathrm{B}}$ from the reference Bayes estimator, based on a prior $\pi_{1.0}$ and symmetric Binder loss $L_{\mathrm{B}}$. 
Analogous to the simulation study, the Bayes estimators are highly similar to each other among the risk equilibrium priors in terms of Binder's distance. 
Also, the estimated number of clusters from Bayes estimators are shown in \cref{table:galaxyclust}. The estimated number of clusters is 3 to 5 with risk equilibrium priors and mostly 3 for risk penalization priors with respect to either Binder's loss or VI risk lower bounds, which generally align with the analysis of \cite{Aitkin2001-zq} who claimed three or four clusters are reasonable. 
Here we emphasize that the estimated number of clusters in \cref{table:galaxyclust} are based on the Bayes estimators, not based on the posterior distribution of the number of clusters. 
The Bayes estimators obtained with VI loss always yield less number of clusters than risk equilibrium priors (with respect to generalized Binder's loss), where the results under the prior $\pi_{0.7}, \dots,\pi_{1.0}$ can be explained with prior risk's behavior based on \cref{thm:gbloss_REprior} and \cref{thm:viloss_RPprior}, supporting our main research question Q2 in Figure~\ref{fig:keyq}(b).

\section{Concluding remarks}

We proposed the notion of risk equilibrium and risk penalization priors associated with a loss function for Bayesian model selection problems on a structured model space. Motivated by the model selection procedure with Bayes estimators, we define the risk equilibrium and penalization priors through the prior risk function's behavior. The risk equilibrium prior can be thought of as a family of objective prior associated with a loss function, which acts as a pivot that calibrates model selection results across different choices of loss function within the same subclass. The comparison with risk penalization priors provides a better understanding of the loss functions' effect on model selection results. We also provide several concrete examples of risk equilibrium priors under Binder's loss for Bayesian cluster analysis. 

The proposed conceptual framework is best suited for model selection problem settings, and the extension of risk equilibrium and penalization priors in domains other than model space $\calM$ is highly nontrivial. To see this, the prior distribution must be proper to be a prior risk well-defined. This is obvious for finite model space $\calM$, but not for continuous domain. One can show that for a 1-dimensional interval domain $[-1,1]$ with absolute and squared error loss, there is no risk equilibrium prior with respect to squared error loss that is proper, and the only proper risk equilibrium prior with respect to absolute error loss is the point masses at $-1$ and $1$ with equal weights. Nevertheless, we hope there are potential alternatives to prior risk functions for domains other than model space, which can be interpreted as scoring rules \citep{Leisen2020-sd}.

Another interesting avenue of future research is the other direction, finding an ``objective'' loss function given a prior distribution. This would yield a family of loss functions that give prior risk to be a constant function. The family of loss function may need to be restricted or parameterized, in such cases \cref{thm:ghloss_REprior} and \cref{thm:gbloss_REprior} could give a partial answer. The characterization of such a family of loss functions would help find new loss functions with interesting properties.

\bibliography{ref_paperpile.bib}
\bibliographystyle{apalike}

\clearpage

\appendix

\section{Appendix}

\subsection{Proofs of Propositions and Theorems}
\label{proof:proofs}
\subsubsection{Proof of Proposition 1}

\begin{proof}
The posterior risk with respect to GH(a) loss is
\begin{align*}
        R^\pi_{\mathrm{GH}(a)}(\hat{\bm\gamma}) &= \sum_{i=1}^p ( a \bbP^\Pi(\gamma_i = 0) \ind(\hat{\gamma}_i = 1) + (2-a) \bbP^\Pi(\gamma_i = 1) \ind(\hat{\gamma}_i = 0) )
\end{align*}
Since $\bbP^\Pi(\gamma_i = 1) > 0.5a$ is equivalent to $a\bbP^\Pi(\gamma_i = 0) < (2-a)\bbP^\Pi(\gamma_i = 1)$ so that choosing $\hat{\gamma}_i = 1$ has a lower risk than $\hat{\gamma}_i = 0$, the minimizer chooses $i$th variable if its marginal inclusion probability is greater than $0.5a$, and otherwise doesn't choose $i$th variable. This proves \cref{prop:ghloss_riskeq}, the $0.5a$ quantile probability model is the Bayes estimator under the GH(a) loss.
\end{proof}   

\subsubsection{Proof of Proposition 2}
\begin{proof}
    The posterior risk with respect to GB(a) loss is
    \begin{align}
R_{\mathrm{GB}(a)}^\Pi(\hat{\bfz}) &= \sum_{i<j}\left(a C_{ij} \ind(\hat{z}_i \neq \hat{z}_j) + (2-a)(1-C_{ij})\ind(\hat{z}_i = \hat{z}_j)\right) \label{eq:gbrisk1} \\
&= \sum_{i<j} aC_{ij} + (2-a-2C_{ij})\ind(\hat{z}_i = \hat{z}_j) \label{eq:gbrisk2}\\
&= \sum_{i<j} \rho_{(2-a)/2}\left(\ind(\hat{z}_i = \hat{z}_j)-C_{ij}\right) \label{eq:gbrisk3}
\end{align}
which proves \cref{prop:gbloss_riskeq}. The last equation can be easily checked using the fact that $\ind(\hat{z}_i =\hat{z}_j)$ is binary and $C_{ij}\in[0,1]$.
\end{proof}

\subsubsection{Proof of Proposition 3.1}

\begin{proof}
    Let $\pi_1,\pi_2\in\calF_\delta$ which are point masses at $M_1$ and $M_2$ respectively. Then $S(\pi_1,\hat{M}) = R^{\pi_1}(\hat{M}) = L(M_1,\hat{M})$ and $S(\pi_1,\pi_2) = L(M_1,M_2)$. By the identity of indiscernibles, $S(\pi_1,\pi_2)$ is minimized at 0 if and only if $M_1=M_2$, which is equivalent to $\pi_1 = \pi_2$, thus $S$ is a strictly proper scoring rule relative to $\calF_\delta$. This proves \cref{prop:scoringrule}.
\end{proof}

\subsubsection{Proof of Theorem 3.2}

\begin{proof} The prior risk with respect to GH(a) loss is
\begin{align*}
        R^\pi_{\mathrm{GH}(a)}(\hat{\bm\gamma}) &= \sum_{i=1}^p ( a \bbP^\pi(\gamma_i = 0) \ind(\hat{\gamma}_i = 1) + (2-a) \bbP^\pi(\gamma_i = 1) \ind(\hat{\gamma}_i = 0) )
\end{align*}
    When $\bbP^\pi(\gamma_i = 1) = 0.5a$ for all $i=1,\dots,p$, the risk becomes constant in terms of $\hat{\bm\gamma}$. Also, this is the only possible scenario. To see this, suppose $\bbP^\pi(\gamma_{i^\star} = 1) \neq 0.5a$ for some $i^\star$. If we consider $\hat{\bm\gamma}_1$ and $\hat{\bm\gamma}_2$ that only differs by $i^\star$th component, then  $R_{H}^\pi(\hat{\bm\gamma}_1) \neq R_{H}^\pi(\hat{\bm\gamma}_2)$, which leads to non-constant prior risk, which proves \cref{thm:ghloss_REprior}.
\end{proof}

\subsubsection{Proof of Theorem 3.3}

\begin{proof} Denote $\bbP^\pi(z_i=z_{j}) = c$ for simplicity, which does not depends on $(i,j)$ since $\pi$ is exchangeable. Under the generalized Binder's loss \eqref{eq:gbinderloss}, the prior risk becomes (see \eqref{eq:gbrisk2})
\begin{equation}
R_{\mathrm{GB}(a)}^\pi(\hat{\bfz}) =  ac\binom{p}{2} + (2-a-2c) \sum_{i<j}\ind(\hat{z}_i = \hat{z}_j)
\end{equation}
It is clear that $R_{\mathrm{GB}(a)}^\pi(\hat{\bfz})$ becomes a constant function when $c=(2-a)/2$. To show other direction where $c=(2-a)/2$ is a necessary condition for an exchangeable prior $\pi$ to be a risk equilibrium prior with respect to $L_{\mathrm{GB}(a)}$, suppose $c\neq (2-a)/2$. Then there are $\hat{\bfz}_1$ and $\hat{\bfz}_2$ such that $R_{B}^\pi(\hat{\bfz}_1) \neq R_{B}^\pi(\hat{\bfz}_2)$; for example $\hat{\bfz}_1$ being a partition of singletons and $\hat{\bfz}_1$ being a one-block partition. This proves \cref{thm:gbloss_REprior}.
\end{proof}

\subsubsection{Proof of Theorem 4.1}

\begin{proof} 
The proof is highly similar to the proof of \cref{thm:ghloss_REprior}. Again, the prior risk with respect to GH(a) loss is
\begin{align*}
        R^\pi_{\mathrm{GH}(a)}(\hat{\bm\gamma}) &= \sum_{i=1}^p ( a \bbP^\pi(\gamma_i = 0) \ind(\hat{\gamma}_i = 1) + (2-a) \bbP^\pi(\gamma_i = 1) \ind(\hat{\gamma}_i = 0) )
\end{align*}
    To see $\bbP^\pi(\gamma_i = 1) \le 0.5a$ for $i=1,\dots,p$ implies risk penalization, by transitivity, it is sufficient to show that prior risk is increasing for the covering pairs $(\hat{\bm\gamma}_1, \hat{\bm\gamma}_2)$, i.e. $\hat{\bm\gamma}_1\prec \hat{\bm\gamma}_2$ and there is no other element between. Let $i^*$ be the index of the added variable, which can be any indices. Then, we have 
    \[
    R^\pi_{\mathrm{GH}(a)}(\hat{\bm\gamma}_2) - R^\pi_{\mathrm{GH}(a)}(\hat{\bm\gamma}_1) = a \bbP^\pi(\gamma_{i^*} = 0) - (2-a) \bbP^\pi(\gamma_{i^*} = 1) \ge 0 
    \]
    where the last inequality comes from  $\bbP^\pi(\gamma_i = 1) \le 0.5a$ for $i=1,\dots,p$. 
    Now we show another direction, suppose that there is an index $i^\star$ such that $\bbP^\pi(\gamma_{i^\star} = 1) > 0.5a$. Consider a pair $\hat{\bm\gamma}_3\prec \hat{\bm\gamma}_4$ such that they only differ in $i^\star$th index and $\hat{\gamma}_{4,i^\star} = 1$. This pair is the one that leads to the prior risk strictly decreasing
     \[
    R^\pi_{\mathrm{GH}(a)}(\hat{\bm\gamma}_4) - R^\pi_{\mathrm{GH}(a)}(\hat{\bm\gamma}_3) = a \bbP^\pi(\gamma_{i^\star} = 0) - (2-a) \bbP^\pi(\gamma_{i^\star} = 1) < 0 
    \]
    which proves the other direction and thus \cref{thm:ghloss_RPprior}.
\end{proof}

\subsubsection{Proof of Theorem 4.2}

\begin{proof}The proof is highly similar to the proof of \cref{thm:gbloss_REprior}. Again, denote $\bbP^\pi(z_i=z_{j}) = c$ for simplicity, which does not depends on $(i,j)$ since $\pi$ is exchangeable. Recall that the prior risk under the generalized Binder's loss \eqref{eq:gbinderloss} is 
\begin{equation}
R_{\mathrm{GB}(a)}^\pi(\hat{\bfz}) =  ac\binom{p}{2} + (2-a-2c) \sum_{i<j}\ind(\hat{z}_i = \hat{z}_j)
\end{equation}
It is clear that when $c \ge (2-a)/2$, 
$R_{\mathrm{GB}(a)}^\pi(\hat{\bfz})$ is a decreasing function of $\sum_{i<j}\ind(\hat{z}_i = \hat{z}_j)$. 

To show that $\pi$ is a risk penalization prior with respect to $\mathrm{GB}(a)$ loss, by transitivity, it is sufficient to show that prior risk is increasing for the covering pairs $(\hat{\bfz}_1, \hat{\bfz}_2)$. If $\hat{\bfz}_1 \prec \hat{\bfz}_2$, which means $\hat{\bfz}_2$ is obtained by splitting one of the cluster of $\hat{\bfz}_1$, we have increasing prior risk $R^\pi_{GB(a)}(\hat{\bfz}_1) \le R^\pi_{GB(a)}(\hat{\bfz}_2)$ since the term $\sum_{i<j}\ind(\hat{z}_i = \hat{z}_j)$ is decreasing by splitting the cluster. To see other direction, suppose $c<(2-a)/2$. Then $R_{\mathrm{GB}(a)}^\pi(\hat{\bfz})$ is an increasing function of $\sum_{i<j}\ind(\hat{z}_i = \hat{z}_j)$, and there is a pair $(\hat{\bfz}_3, \hat{\bfz}_4)$ (actually, all pairs) such that $\hat{\bfz}_3\prec \hat{\bfz}_4$ but prior risk is strictly decreasing. This proves \cref{thm:gbloss_RPprior}.
\end{proof}

\subsubsection{Proof of Theorem 4.3}

\begin{proof} Let $c = \bbP^\pi(z_i = z_j)$ be a prior co-clustering probability for $i\neq j$ under the exchangeable prior $\pi$. The lower bound of prior VI risk can be written as (compare with \eqref{eq:virisklb})
\begin{equation}
\tilde{R}_V^\pi(\hat{\bfz}) = h + \frac{1}{p}\sum_{i=1}^p \log_2 \left(\sum_{j=1}^p \ind(\hat{z}_i=\hat{z}_j)\right) - \frac{2}{p}\sum_{i=1}^p\log_2 \left(1+c\sum_{j\neq i} \ind(\hat{z}_i=\hat{z}_j)\right) 
\end{equation}
where $h = \bbE^\pi\left[p^{-1}\sum_{i=1}^p \log_2 \sum_{j=1}^p \ind(z_i=z_j)\right]$ is a constant. Multiplying both sides by $p$ and changing the sum over the data indices to sum over the cluster indices $l =1,\dots,k$ where $k$ is the number of clusters in $\hat{\bfz}$,
\begin{align}
 p\tilde{R}_V^\pi(\hat{\bfz}) &=  ph + \sum_{i=1}^p   \log_2 \frac{\sum_{i=1}^p 1(\hat{z}_i = \hat{z}_j)}{(1+ \rho\sum_{j\neq i} 1(\hat{z}_i = \hat{z}_j) )^2} \\
 &= ph +  \sum_{l=1}^k \left\{n_l\log_2\left(\frac{n_l}{(1+c (n_l-1))^2}\right)\right\} ,\label{eq:virisklb2}
\end{align}
here $n_l$ is the size of the $l$th cluster of $\hat\bfz$. 

We claim that function $g(m) = m \log_2(m/(1+c(m-1))^2)$ defined on domain $m\in\bbN$ is subadditive when $c \ge \sqrt{2}-1$, i.e.  $g(m_1 + m_2) \le g(m_1) + g(m_2)$ for any $m_1,m_2 \in \bbN$.  Now let $\hat{\bfz}'$ be a splitted cluster from $\hat{\bfz}$. Then, by subadditivity of $g(m)$ on $\bbN$ and equation \eqref{eq:virisklb2}, we have $ \tilde{R}_V^\pi(\hat{\bfz}) \le \tilde{R}_V^\pi(\hat{\bfz}')$.
To see the claim, it can be easily shown that $g(m)/m = \log_2(m/(1+c(m-1))^2)$ is decreasing on $m\in\bbN$ when $c \ge \sqrt{2}-1$. Combining $m_1g(m_1+m_2)/(m_1+m_2) \le g(m_1)$ and $m_2g(m_1+m_2)/(m_1+m_2) \le g(m_2)$, we have $g(m_1 + m_2) \le g(m_1) + g(m_2)$. 

To show the other direction, suppose $c < \sqrt{2}-1$, which implies $g(2) > 2g(1)$. Then, let $\hat{\bfz}_{3}$ and $\hat{\bfz}_4$ be two partitions where $\hat{\bfz}_4$ is obtained by splitting one of the cluster of $\hat{\bfz}_3$ with size 2 to singleton clusters. Then we have $\hat\bfz_3\prec \hat\bfz_4$ but $\tilde{R}_V^\pi(\hat\bfz_3) > \tilde{R}_V^\pi(\hat\bfz_4)$ since $g(2) >2g(1)$, which shows $\pi$ is not a risk penalization prior with respect to $\tilde{R}_V$ and proves \cref{thm:viloss_RPprior}.
\end{proof}

\subsubsection{Details of Example 4.2}
\label{appendix:yangRPproof}

We show that truncated prior $\pi(\bm\gamma) \propto p^{-\kappa |\bm\gamma|}\ind( |\bm\gamma|\le s_{\mathrm{max}})$ of \cite{Yang2016-yu} is a risk penalization prior with respect to $L_{\mathrm{H}}$.

\begin{proof}
    To show that $\pi$ is a risk penalization prior with respect to Hamming loss, it is sufficient to show that the prior marginal inclusion probability is less than 0.5 for all variables. The prior $\pi(\bm\gamma) \propto p^{-\kappa |\bm\gamma|}\ind(|\bm\gamma|\le s_{\mathrm{max}})$ with $\kappa \ge 2$ that \cite{Yang2016-yu} considered can be thought of as hierarchical prior based on model size, where $\pi(\bm\gamma) = \pi(\bm\gamma \given s)q(s)$ with $\pi(\bm\gamma\given s) = \binom{p}{s}^{-1}$ and $q(s) \propto \binom{p}{s}p^{-\kappa s} \ind(s\le s_{\mathrm{max}})$. By symmetry, the prior marginal inclusion probability for any variable $i$ are all the same and can be written as, using the hierarchical representation,
    \begin{equation}
    \label{eq:yangpriorinclusionprob}
    \bbP(\gamma_i = 1) = \sum_{s=0}^{s_{\mathrm{max}}}\frac{\binom{p-1}{s-1}}{\binom{p}{s}}q(s) = \sum_{s=0}^{s_{\mathrm{max}}} \frac{s}{p} \frac{\binom{p}{s} p^{-\kappa s}}{\sum_{s'=0}^{s_{\mathrm{max}}}\binom{p}{s'}p^{-\kappa s'}}    
    \end{equation}
    
We claim that $\frac{\binom{p}{s} p^{-\kappa s}}{\sum_{s'=0}^{s_{\mathrm{max}}}\binom{p}{s'}p^{-\kappa s'}}\le \frac{1}{p+1}$ for any given $s\ge 1$ and $s_{\mathrm{max}}\ge 1$. To see this, denominator is greater than $1+p^{-1}$ and numerator is less than $p^{-1}$, thus it has upper bound $(p+1)^{-1}$ that does not depend on $s$ and $s_{\mathrm{max}}$. Therefore, we have an upper bound of \eqref{eq:yangpriorinclusionprob}, $\sum_{s=0}^p(s/p)(1/(p+1)) = 1/2$ by setting $s_{\mathrm{max}} =p$, which proves $\pi$ is risk penalization prior with respect to Hamming loss.
\end{proof}

\newpage

\subsection{Counterexamples on risk equilibrium and risk penalization priors}
\label{appendix:counterexamples}

Throughout this subsection, we consider a BVS model space $\calM = \{0,1\}^2$ with $p=2$, denoted as $\calM = \{M_{00}, M_{01}, M_{10}, M_{11}\}$ and satisfies $M_{00}\prec M_{01} \prec M_{11}$ and $M_{00}\prec M_{10} \prec M_{11}$. We simply denote prior probabilities by a 4-tuple $\pi = (\pi_{00}, \pi_{01}, \pi_{10}, \pi_{11})$ and similarly for prior risks $R^\pi = (R^\pi(M_{00}), R^\pi(M_{01}), R^\pi(M_{10}), R^\pi(M_{11})) $. We consider two loss functions, denoted as $L_1$ and $L_2$, defined as follows.

    \begin{center}
    \vspace{2mm}
\begin{tabular}{c c c c c}
    \toprule
  $L_1$ & $M_{00}$ & $M_{01}$  & $M_{10}$ & $M_{11}$ \\
    \midrule
   $M_{00}$ & 0 & 1& 1& 2 \\
    \midrule
  $M_{01}$ & 1& 0 & 3 & 4\\
  \midrule
  $M_{10}$ & 1& 3 & 0& 4\\
  \midrule
  $M_{11}$ & 2& 4& 4& 0\\
   \bottomrule
\end{tabular}
\hspace{5mm}
\begin{tabular}{c c c c c}
    \toprule
  $L_2$ & $M_{00}$ & $M_{01}$  & $M_{10}$ & $M_{11}$ \\
    \midrule
   $M_{00}$ & 0 & 1& 1& 3 \\
    \midrule
  $M_{01}$ & 1& 0 & 1 & 2\\
  \midrule
  $M_{10}$ & 1& 1 & 0& 2\\
  \midrule
  $M_{11}$ & 3& 2& 2& 0\\
   \bottomrule
\end{tabular}
\vspace{2mm}
\end{center}

Those two loss functions are both symmetric and satisfy the identity of indiscernibles, but those are pathological examples in the sense that they do not ``conform'' to the geometry of the model space or have a peculiar asymmetry. For example, we have $2 = L_1(M_{00}, M_{11}) < L_1(M_{01},M_{11}) = 4$, which does not conform with the ordering $M_{00} \prec M_{01} \prec M_{11}$. In the second example, we have $L_2(M_{00},M_{11}) = 3$ but $L_2(M_{01},M_{10}) = 1$, showing the loss is not invariant of zero/one labels.

\begin{example}[Risk equilibrium prior may not exist] Under the loss $L_1$, to find the risk equilibrium prior, we solve the following systems of equations along with equation $\pi_{00} + \pi_{01} + \pi_{10} + \pi_{11} = 1$,
    \[
R^{\pi}(M_{00}) \stackrel{(i)}{=} R^{\pi}(M_{01}) \stackrel{(ii)}{=} R^{\pi}(M_{10}) = R^{\pi}(M_{11})
\]
First, equation $(ii)$ leads to $\pi_{01} = \pi_{10}$, and then equation $(i)$ gives $\pi_{00} + \pi_{01} + 2\pi_{11} = 0$, which implies $\pi_{00} = \pi_{01} = \pi_{10} = \pi_{11} = 0$, a contradiction. Thus, risk equilibrium prior with respect to $L_1$ does not exist.
\end{example}

\begin{example}[Risk equilibrium prior may not have a full support]
    Suppose there is a risk equilibrium prior $\pi = (\pi_{00}, \pi_{01}, \pi_{10}, \pi_{11})$ with respect to loss $L_2$. Then, solving the linear system of equations by equating the prior risks yields the risk equilibrium prior $\pi = (1/2, 0, 0, 1/2)$, assigning zero probability to $M_{01}$ and $M_{10}$.
\end{example}

\begin{example}[Risk penalization priors does not imply decreasing prior probabilities]
Consider a prior $\pi = (0.2, 0.25, 0.25, 0.3)$ which has increasing prior probabilities in terms of model complexity. But at the same time, $\pi$ is a risk penalization prior with respect to loss function $L_1$ since the prior risk becomes $R^\pi_1 = (1.1, 2.15, 2.15, 2.4)$.
\end{example}

\begin{example}[Decreasing prior probabilities does not imply risk penalization priors]
Consider a prior $\pi = (0.3, 0.25, 0.25, 0.2)$ which has decreasing prior probabilities in terms of model complexity. The prior risk with respect to loss function $L_2$ has prior risks $R^\pi_2 = (1.1, 0.95, 0.95, 1.9)$, which is not increasing in terms of model complexity and thus $\pi$ is not a risk penalization prior.
\end{example}

\subsection{Additional table}
\label{appendix:moretable}

\begin{table}[htbp]
\small
    \centering
   \caption{Average model size $|\bm\gamma|$ of Bayes estimators in BVS simulations with three different $n$. $L_{\mathrm{GH}(a)}$ denotes generalized Hamming loss with weight $a$ and $L_{01}$ denotes zero-one loss. Underlines indicate results from risk equilibrium priors. The lower diagonal part with loss $L_{\mathrm{GH}(a)}$ corresponds to risk penalization priors.}
   \label{table:bvssim2}
    \begin{tabular}{c cccccccc c}
    \toprule
            &  \multirow{2}{*}{Loss}  & \multicolumn{8}{c}{Model priors ($\pi_a \equiv$  Beta-Binomial$(a, 2-a)$)}  \\
      &    & $\pi_{0.7}$ & $\pi_{0.8}$ & $\pi_{0.9}$ & $\pi_{1.0}$ &$\pi_{1.1}$ & $\pi_{1.2}$ & $\pi_{1.3}$ & $\pi_{\mathrm{unif}}$\\
        \midrule
    \multirow{8}{*}{$n=20$}  & $L_{\mathrm{GH}(0.7)}$ &  \underline{2.861} & 3.420 & 4.065 & 4.791 & 5.581 & 6.423 & 7.382 & 8.680 \\ 
      &  $L_{\mathrm{GH}(0.8)}$ & 2.323 & \underline{2.812} & 3.376 & 3.988 & 4.720 & 5.483 & 6.362 & 6.819 \\ 
      & $L_{\mathrm{GH}(0.9)}$ & 1.887 & 2.323 & \underline{2.801} & 3.313 & 3.967 & 4.697 & 5.469 & 5.536 \\ 
      & $L_{\mathrm{GH}(1.0)}$ & 1.542 & 1.919 & 2.310 & \underline{2.780} & 3.289 & 3.917 & 4.701 & \underline{4.621} \\ 
      &  $L_{\mathrm{GH}(1.1)}$ & 1.283 & 1.566 & 1.936 & 2.294 & \underline{2.757} & 3.299 & 3.949 & 3.884 \\ 
      & $L_{\mathrm{GH}(1.2)}$ & 1.081 & 1.302 & 1.584 & 1.905 & 2.282 & \underline{2.732} & 3.288 & 3.244 \\ 
      &  $L_{\mathrm{GH}(1.3)}$ & 0.870 & 1.077 & 1.311 & 1.571 & 1.877 & 2.240 & \underline{2.726} & 2.701 \\ 
      \cmidrule{2-10}
      &  $L_{01}$ & 0.646 & 1.084 & 1.471 & 1.951 & 2.830 & 3.745 & 4.694 & \underline{4.613} \\ 
        \midrule
    \multirow{7}{*}{$n=60$}  & $L_{\mathrm{GH}(0.7)}$ & \underline{5.362} & 5.800 & 6.220 & 6.689 & 7.245 & 7.858 & 8.558 & 7.007 \\ 
      & $L_{\mathrm{GH}(0.8)}$ & 4.762 & \underline{5.086} & 5.468 & 5.889 & 6.376 & 6.838 & 7.494 & 6.306 \\ 
      & $L_{\mathrm{GH}(0.9)}$& 4.262 & 4.582 & \underline{4.874} & 5.234 & 5.597 & 6.087 & 6.611 & 5.721 \\ 
      & $L_{\mathrm{GH}(1.0)}$ & 3.811 & 4.082 & 4.388 & \underline{4.684} & 5.015 & 5.399 & 5.877 & \underline{5.245} \\ 
      & $L_{\mathrm{GH}(1.1)}$ & 3.446 & 3.670 & 3.929 & 4.196 & \underline{4.492} & 4.832 & 5.215 & 4.823 \\ 
      & $L_{\mathrm{GH}(1.2)}$ &  3.117 & 3.309 & 3.530 & 3.759 & 4.018 & \underline{4.313} & 4.648 & 4.440 \\ 
      & $L_{\mathrm{GH}(1.3)}$&  2.804 & 3.001 & 3.188 & 3.380 & 3.603 & 3.856 & \underline{4.128} & 4.085 \\ 
      \cmidrule{2-10}
      &  $L_{01}$ & 2.962 & 3.391 & 3.959 & 4.566 & 5.344 & 6.309 & 7.401 & \underline{5.066} \\ 
        \midrule
          \multirow{7}{*}{$n=1000$}  & $L_{\mathrm{GH}(0.7)}$ & \underline{6.361} & 6.371 & 6.386 & 6.406 & 6.419 & 6.449 & 6.475 & 6.345 \\
      &  $L_{\mathrm{GH}(0.8)}$ & 6.290 & \underline{6.303} & 6.311 & 6.323 & 6.337 & 6.353 & 6.373 & 6.265 \\
      &  $L_{\mathrm{GH}(0.9)}$ & 6.213 & 6.229 & \underline{6.239} & 6.253 & 6.264 & 6.283 & 6.295 & 6.203 \\ 
      &  $L_{\mathrm{GH}(1.0)}$ & 6.170 & 6.177 & 6.190 & \underline{6.198} & 6.204 & 6.212 & 6.226 & \underline{6.153} \\ 
      &  $L_{\mathrm{GH}(1.1)}$ & 6.127 & 6.135 & 6.141 & 6.149 & \underline{6.158} & 6.166 & 6.181 & 6.126 \\ 
      &  $L_{\mathrm{GH}(1.2)}$ & 6.104 & 6.109 & 6.112 & 6.113 & 6.120 & \underline{6.127} & 6.132 & 6.101 \\ 
      &  $L_{\mathrm{GH}(1.3)}$ & 6.086 & 6.090 & 6.093 & 6.093 & 6.097 & 6.104 & \underline{6.109} & 6.083 \\ 
        \cmidrule{2-10}
      &  $L_{01}$ &  6.117 & 6.121 & 6.134 & 6.138 & 6.142 & 6.147 & 6.152 & \underline{6.147} \\ 
      \bottomrule
    \end{tabular}
\end{table}

\end{document}